\documentstyle[preprint,seceq,epsf,wrapft]{ptptex}

\notypesetlogo  
\preprintnumber[3cm]{
KUNS-1502\\ HE(TH)~98/04\\ hep-th/9804114}

\title{
Comments on Duality in MQCD
}

\author{
Shigeki {\sc Sugimoto}\footnote{E-mail:
  sugimoto@gauge.scphys.kyoto-u.ac.jp}$^,$
\footnote{Supported in part by the Grant-in-Aid
for JSPS fellows.}
}

\inst{
Department of Physics, Kyoto University,\\
Kyoto 606-8502, Japan
}

\abst{
We clarify some ambiguous points in a derivation of duality via
 brane exchange using M-theory language,
and propose a ``proof'' of duality in MQCD.
Actually, duality in MQCD is rather trivial and 
does not need a complicated proof.
 The problem is how to interpret it in  field
theory language.
We examine BPS states in $\N=2$ theory and find
the particle correspondence under duality.
In the process, we also find some exotic particles in $\N=2$ MQCD,
and we observe an interesting phenomenon in type IIA string theory,
namely, that fundamental strings are converted into D2-branes via the
exchange of two NS5-branes.
We also discuss how we should understand Seiberg's $\N=1$ duality
from exact duality in MQCD.
}

\begin{document}
\font\mybb=msbm10 at 12pt
\def\bb#1{\hbox{\mybb#1}}
\def\Z {\bb{Z}}
\def\R {\bb{R}}
\def\C {\bb{C}}
\def\J {\bb{J}}
\def\I {\bb{I}}
\def\CP {\bb{P}}

\font\mycc=msbm10 at 8pt
\def\cc#1{\hbox{\mycc#1}}
\def\sZ {\cc{Z}}
\def\sR {\cc{R}}
\def\sC {\cc{C}}
\def\sJ {\cc{J}}
\def\sI {\cc{I}}
\def\sCP {\cc{P}}

\def\ele{\mathop{\rm ele}\nolimits}
\def\mag{\mathop{\rm mag}\nolimits}
\def\tr{\mathop{\rm tr}\nolimits}
\def\Im{\mathop{\rm Im}\nolimits}
\def\diag{\mathop{\rm diag}\nolimits}
\def\rank{\mathop{\rm rank}\nolimits}
\def\Tr{\mathop{\rm Tr}\nolimits}
\def\mod{\mathop{\rm mod}\nolimits}
\def\real{\mathop{\rm Re}\nolimits}

\renewcommand{\thefootnote}{\fnsymbol{footnote}}
\renewcommand{\theequation}{\thesection.\arabic{equation}}
\makeatletter
\@addtoreset{equation}{section}
\@addtoreset{footnote}{page}
\makeatother

\newcommand{\vs}{\vspace*}
\newcommand{\hs}{\hspace*}
\newcommand{\wt}{\widetilde}
\newcommand{\ol}{\overline}
\newcommand{\ul}{\underline}
\newcommand{\ra}{\rightarrow}
\newcommand{\lra}{\leftrightarrow}
\newcommand{\nn}{\nonumber}
\newcommand{\mms}[1]{\makebox[4ex]{$#1$}}
\newcommand{\sq}{\sqrt{2}\,}
\newcommand{\VEV}[1]{\left\langle #1\right\rangle}
\newcommand{\norm}[1]{\parallel #1\parallel}

\newcommand{\N}{{\cal N}}
\newcommand{\beq}{\begin{eqnarray}}
\newcommand{\eeq}{\end{eqnarray}}

\maketitle

\section{Introduction}

Since Seiberg's duality \cite{S} in $\N=1$ SQCD  appeared, various
examples of
 duality have been found, and several attempts to expose the mysterious
nature of duality have been made. However, there is
still no proof of the
 duality, and we do not really know why and
 when there is a dual description.

One of the most fancy derivations of Seiberg's duality
is obtained via brane exchange in type IIA string 
theory,\cite{EGK,EGKRS,BSTY} extending the idea 
given in Ref.~\citen{HW}.
The authors of that paper
 argued that the electric theory in Seiberg's duality 
(i.e. $\N=1$ $SU(N_c)$ SQCD with $N_f$ flavors)
 can be realized as an effective world volume theory setting
two NS5-branes, $N_c$ D4-branes and $N_f$ D6-branes,
 in a certain configuration.
Then, if one gradually deform the configuration and exchanges
 the two NS5-branes,
the effective theory turns out to be the magnetic theory 
(i.e. $\N=1$ $SU(N_f-N_c)$ SQCD with $N_f$ flavors
 and a gauge singlet meson)!
This argument explains the fact that the vacuum moduli spaces in the
 electric and magnetic theories are the same.
However, since the brane configurations are not the same, we think 
that it is still unclear whether or not the electric and magnetic theories
are really equivalent.

Recently it has become clear that four-dimensional field theory
 can be analyzed via M-theory technology.\cite{W}
The brane configurations for the electric and magnetic theories in type IIA
string theory can be lifted to M5-brane configurations in M-theory.
In the M-theory language, SQCD is realized as an effective field theory on
the M5-brane world volume wrapped on a Riemann surface, which is
often called MQCD.

It is not difficult to lift the above
arguments to M-theory.\cite{SS,CS,F}
 Many results are beautifully reproduced
using M-theory language.
The key ingredient we think is that
the electric and magnetic theories should be understood
as the effective world volume theory of the {\it identical} M5-brane. 
In this context, the electric and magnetic theories are, by definition,
exactly equivalent. Our motivation for this work is to understand
duality in this way and make a dictionary 
translating between the electric and magnetic
descriptions. 
 We wish to emphasize that MQCD is a nice formulation
of gauge theories, which is quite compatible with duality.

In \S \ref{sect2}, we review some field theory
results given in Ref.~\citen{APS,HMS}. In \S \ref{sect3}
we propose a ``proof'' of duality in MQCD, and show that the results
in \S \ref{sect2} can be reproduced consistently.
We also discuss how we should understand
Seiberg's duality from the exact duality in MQCD.
Section \ref{sect4} is devoted to exploring the correspondence of BPS
states under duality in $\N=2$ theory. 
It will become clear that the magnetic theory can be understood
 as the soliton sector of the electric theory.
Interpreting this result in the IIA picture, we observe the interesting
phenomenon that fundamental strings are converted into D2-branes via
the exchange of two NS5-branes.
Here we give some explicit examples of
BPS states, finding holomorphic surfaces
embedded in a multi-Taub-NUT space.
We find W-bosons and quarks, which are the elementary particles
in the field theory, together with many exotic particles,
which can never be obtained in the perturbative field theory.
We also make some comments on the realization of BPS states
via geodesics on the Riemann surface.
In \S \ref{sect5}, we make our conclusion
 and discuss future directions to be pursued.

\section{The field theory approach}
\label{sect2}
\subsection{Duality in $\N=2$ SQCD}
\label{sect2.1}

In this subsection, we review some semi-classical results
in $\N=2$ SQCD given in Ref.~\citen{APS,HMS}.
Here we consider $\N=2$ $SU(N_c)$ SQCD with $2N_c$ flavors. 
The theory can be described in terms of ${\cal N}=1$
superfields: $W_\alpha$ (a field strength chiral superfield), $\Phi$
(a chiral superfield in the adjoint representation
of the gauge group), $Q^i$ and
$\wt{Q}_i$ (chiral superfields in the ${\bf N_c}$ and ${\bf \ol{N_c}}$
representation of the gauge group, respectively),
where $i=1,\cdots,2N_c$ are flavor
indices. The superpotential is
\beq
  W_{\ele}= Q^i\Phi \wt{Q}_i + m_i^j Q^i\wt{Q}_j,
\label{Wele}
\eeq
where $m=(m_i^j)=\diag(m_1,\cdots,m_{2N_c})$ is a quark mass matrix.
The bare gauge coupling constant is denoted as
 $\tau=\frac{8\pi}{g^2}i+\frac{\theta}{\pi}$.
We call this theory the `electric theory'.


The basic holomorphic gauge invariant operators which parameterize the 
vacuum moduli space are meson and baryons
defined as follows:
\begin{eqnarray}
  M^i_j&\equiv&Q^i_a\wt{Q}_j^a, \\
  B^{i_1\cdots i_{N_c}} &\equiv&
  Q^{i_1}_{a_1}\cdots Q^{i_{N_c}}_{a_{N_c}}\epsilon^{a_1\cdots a_{N_c}}, \\
  \wt{B}_{i_1\cdots i_{N_c}} &\equiv&
  \wt{Q}_{i_1}^{a_1}\cdots \wt{Q}_{i_{N_c}}^{a_{N_c}}
  \epsilon_{a_1\cdots a_{N_c}}.
\end{eqnarray}

The dual theory has the same matter content and superpotential
as in the electric theory,
\beq
  W_{\mag}= q_i\varphi \wt q^i + {m'}_j^i q_i\wt q^j,
\label{Wmag}
\eeq
but coupling constants are different. 
The quark mass matrix and gauge coupling constant for the dual theory are
 given as $m'= m-\frac{1}{N_c}\Tr m$ and $\tau'=-1/\tau$, respectively.
We call this theory the `magnetic theory'.
Meson and baryons for the magnetic theory are defined as
\begin{eqnarray}
  N_j^i&\equiv&q_{aj}\wt{q}^{ai}, \\
  b_{i_1\cdots i_{N_c}}&\equiv&q_{a_1i_1}\cdots q_{a_{N_c}i_{N_c}}
  \epsilon^{a_1\cdots a_{N_c}}, \\
  \wt{b}^{i_1\cdots i_{N_c}}&\equiv&
\wt{q}^{a_1i_1}\cdots \wt{q}^{a_{N_c}i_{N_c}}
  \epsilon_{a_1\cdots a_{N_c}}.
\end{eqnarray}

The electric and magnetic theories are conjectured to be equivalent
under the correspondence \cite{LS,APS,HMS}

\begin{eqnarray}
  \mbox{electric} & \lra & \mbox{magnetic}  \nn\\
  M     & \lra &  N' ,          \label{MN}\\
 ( M'   & \lra &  N ), \\
  B     & \lra &  (-1)^{N_c} * b , \\
  \wt{B}& \lra &  * \wt{b},
\end{eqnarray}
where $N'\equiv N-\frac{1}{N_c}\Tr N$, $M'\equiv M-\frac{1}{N_c}\Tr M$,
 and
 ``$*$'' indicates the  contraction of
 all flavor indices with the totally
  antisymmetric tensors
$\epsilon_{i_1\cdots i_{2N_c}}$ or $\epsilon^{i_1\cdots i_{2N_c}}$.

Now we set the quark mass matrix as
\beq
 m&=&\diag (0,\cdots,0,m_{N_f+1},\cdots,m_{2N_c}),
 \label{ma}\\
 m'&\equiv&m-\frac{1}{N_c}\Tr m, \nn\\
 &=&\diag (-2m_S,\cdots,-2m_S,m_{N_f+1}-2m_S,\cdots,m_{2N_c}-2m_S),
\eeq
where we have defined $m_S\equiv\frac{1}{2N_c}\Tr m$ and the 
$m_i$ are chosen
to be generic.

Let us consider the baryonic branch on which $B$ or $\wt B$ is
non-zero. It exists only for the case $N_c\leq N_f$.
It is easy to derive semi-classical vacuum expectation
values of the adjoint fields on the baryonic branch, using classical
 F-term equations \cite{APS,HMS}. The results are
\beq
 \Phi&=&0,
 \label{phi_ele}\\
 \varphi&=&\diag(2m_s,\cdots,2m_s,2m_s-m_{N_f+1},\cdots,2m_s-m_{2N_c}).
 \label{phi_mag}
\eeq

{}From these forms, we see that the unbroken gauge groups of
 the electric and magnetic theories are $SU(N_c)$ and $SU(N_f-N_c)
\times U(1)^{2N_c-N_f}$, respectively.
Note that since there are no symmetries which restrict the $\tau$ dependence
in the F-term equations,
we cannot exclude the possibility that these results are corrected
due to some non-perturbative effects. 
Actually, we will see later, using M-theory language,
that there are such non-perturbative corrections.

\subsection{Duality in $\N=1$ SQCD}
\label{N=1SQCD}
Here we summarize some semi-classical results in $\N=1$
theory given in Ref.~\citen{HMS}.
Let us consider $\N=1$ deformed theory, adding a mass term for
 the adjoint chiral multiplet to the electric theory (\ref{Wele}):
\beq
  W_{\ele}= Q^i\Phi \wt{Q}_i + m_i^j Q^i\wt{Q}_j+\frac{\mu}{2}\tr \Phi^2.
\eeq
It was argued in Ref.~\citen{HMS} that the magnetic theory (\ref{Wmag})
should be deformed as
\beq
  W_{\mag}= q_i\varphi \wt{q}^i + {m'}^i_j q_i\wt{q}^j
-\frac{\mu}{2}\tr \varphi^2.
\eeq
The correspondence of the meson operators is also deformed as
\begin{eqnarray}
  \mbox{electric} & \lra & \mbox{magnetic}  \nn\\
  M     & \lra &  \wt N,  \\
 ( \wt M &\lra & N ),
\end{eqnarray}
where we have defined $\wt N \equiv N'+\mu m'$ and  $\wt M\equiv M'-\mu m$.

We set the quark mass as in (\ref{ma}) again and consider the baryonic branch.
The vacuum expectation values of the adjoint fields
are the same as (\ref{phi_ele}) and (\ref{phi_mag}). The meson VEV is,
up to complexified flavor symmetry transformations,
\beq
M=\left(
\begin{array}{cccc}
\rho\\
&\ddots&\\
&&\rho&\hspace{8ex}\\
\\ \\ \\
\end{array}
\right),
\label{meson_e}
\eeq
\beq
N=\wt M=\left(
\begin{array}{ccccccc}
\\ \\ \\ \\ \\
\hspace{18ex}&-\rho\\
&&\ddots\\
&&&-\sigma_{N_f+1}\\
&&&&\ddots\\
&&&&&-\sigma_{2N_c}\\
\end{array}
\right),
\label{meson_m}
\eeq
 where we have defined $\sigma_i\equiv \rho+\mu m_i$.

We can derive Seiberg's duality from the relation $N=\wt M$.
Let us consider the baryonic root, where we have
 $M=Q\wt Q=0$ and $N=q\wt q=-\mu m$.
Taking the D-term equations into account, we obtain
\beq
Q=\wt Q&=&0,\\
q=\wt q&=&\left( 
  \begin{array}{ccccccccccc}
     \\ \\ \\
     
     &\hs{10ex}&&\kappa_{N_f+1}\\
     &&&&\ddots  \\
     &&&&&\kappa_{2N_c}
  \end{array}
\right)
  \begin{array}{l}
     \uparrow\\ 
     N_f-N_c \\
     \downarrow \\
     \uparrow\\ 
     2N_c-N_f \\
     \downarrow \\
  \end{array},
\label{q}
\eeq
where $\kappa_i\equiv\sqrt{-\mu m_i}$.
{}From these forms we expect that the $SU(N_c)$ gauge symmetry 
is unbroken in the electric theory, while $SU(N_c)$ is broken to
$SU(N_f-N_c)$ in the magnetic theory. 

Of course, there are many ambiguities in the semi-classical arguments
given above. In the following sections, we will re-derive these results 
in the M-theory language, which is reliable even in the strong coupling
regime, and try to clarify how to understand duality in this context.

\section{A ``Proof'' of duality in MQCD}
\label{sect3}
\subsection{Duality in $\N=2$ MQCD}

As in Ref.~\citen{W}, we define $\N=2$ MQCD as
an effective world volume theory on an M5-brane wrapped on
a Riemann surface $\Sigma$
embedded holomorphically in
$\C\times\C^*=\{~(v,t)~| ~v=x^4+x^5i,~t=e^{-s},~s=x^6+x^{10}i\,\}$.
In order to obtain the electric theory in \S \ref{sect2.1}
(i.e. $\N=2$ $SU(N_c)$ MQCD with $2N_c$ flavors),
we choose $\Sigma$ to
be the Seiberg-Witten curve given in Ref.~\citen{APSh1}
\begin{eqnarray}
  t^2-2\prod^{N_c}_{a=1}(v+\phi_a)\, t
  +(1-h(\tau)^2)\prod^{2N_c}_{j=1}\left(v-m_j+(1-h(\tau))m_S\right)=0, 
\label{ele}
\end{eqnarray}
where we have defined $h(\tau)\equiv
\frac{\vartheta_3(\tau)^4}{\vartheta_4(\tau)^4-\vartheta_2(\tau)^4}$.

Similarly, we can define the magnetic theory by choosing $\Sigma$
to be
\begin{eqnarray}
  t^2-2\prod^{N_c}_{a=1}(v+\phi_a)\, t
  +(1-h(\tau')^2)\prod^{2N_c}_{j=1}\left(v-m'_j+(1-h(\tau'))m'_S\right)=0, 
\label{mag} 
\end{eqnarray}
where  $\tau'=-1/\tau$, $m'=m-\frac{1}{N_c}\Tr m$ and
 $m'_S=\frac{1}{2N_c}\Tr m'$, which are the gauge coupling constant
and the quark mass parameters for the magnetic theory
suggested in \S \ref{sect2.1}.

Using the relation $h(-1/\tau)=-h(\tau)$,
it is easy to see that (\ref{ele}) and (\ref{mag})
 are the same\cite{APSh1}.
Hence the electric and magnetic theories in MQCD are exactly
equivalent, by definition.
The duality transformation is merely a change of description.
It may be suitable to refer to the electric and magnetic theories as
the electric and magnetic description of MQCD, respectively.

Let us make some comments on the brane exchange in IIA picture.
In the electric description, the weak coupling region is $\tau\sim i\infty$.
The asymptotic positions of the two NS5-branes in the $s=x^6+x^{10}i$ plane 
can be read from the value of $t=e^{-s}$ at $v\ra\infty$, 
and we obtain \cite{NOYY}
\beq
\delta s&\equiv& s|_{NS5(1)}-s|_{NS5(2)},\\
&=&\log(1+h)-\log(1-h),
\label{del_s}\\
&\sim&-i\pi\tau.~~~~~~~~~(\tau\sim i\infty)
\eeq
If we fix the quark mass parameters while moving
 $\tau$ to the strong coupling region, and set $\tau\sim 0$,
we can see from (\ref{del_s}) that the two NS5-branes are exchanged
as argued in Ref.~\citen{NOYY}.
In this region, the magnetic description is weak coupling.
The brane exchange changes the weak coupling electric theory
into the strong coupling electric theory,
which is equivalent to the weak coupling magnetic theory.
Notice that the brane exchange does {\it not} change one theory into
another equivalent theory,
since the weak coupling electric theory and the weak coupling magnetic
theory are, in general, not equivalent.
For example, when we use $m_S\ne 0$, the Riemann surface
(\ref{ele}) is not invariant under brane exchange,
and the effective action computed as in Ref.~\citen{SW}
is explicitly
deformed via brane exchange.
If we tune $m_S$ to  zero, the brane exchange takes one theory
to the same theory. However, it is not worth referring
to this as `duality',
since the brane exchanged theory has the same matter content and
couplings as the original one. 
We emphasize that we refer to the exact equivalence of
 the weak (strong) coupling electric theories and
the strong (weak) coupling magnetic theories (respectively) as `duality'.
This is highly non-trivial in usual field theory, but trivial in
MQCD.

In order to compare the results in MQCD with the semi-classical
results given in the previous section, we should analyze
the weak coupling region. Let us check that the
results given in the last section can be reproduced in the weak
coupling limits in the electric and magnetic descriptions
of MQCD.

Now we set the quark mass matrix as in (\ref{ma}) and consider
the baryonic branch root as considered in Ref.~\citen{APS,NOYY,F}.
 On the baryonic branch, the curve (\ref{ele}) is 
factorized \cite{APS,HOO} as
\beq
(t-\alpha(v))(t-\beta(v))=0,
\label{fact}
\eeq
where $\alpha(v)$ and $\beta(v)$ are some polynomials satisfying
\beq
\alpha(v)+\beta(v)&=&2\prod_{a=1}^{N_c}(v+\phi_a),\\
\alpha(v)\beta(v)&=&(1-h^2)\prod^{2N_c}_{j=1}(v-m_j+(1-h)m_S).
\eeq
The solution exists iff the $\phi_a$ satisfy
\beq
2\prod_{a=1}^{N_c}(v+\phi_a)=(1+h)\,\hat v^{N_c}
+(1-h)\,\hat v^{N_f-N_c}\prod^{2N_c}_{j=N_f+1}(\hat v-m_j),
\label{phi_M}
\eeq
where we have defined $\hat v\equiv v+(1-h)m_S$.
Then the solution is
\beq
\alpha(v)&=&(1+h)\,\hat v^{N_c},
\label{alpha}\\
\beta(v)&=&(1-h)\,\hat v^{N_f-N_c}\prod^{2N_c}_{j=N_f+1}(\hat v-m_j).
\label{beta}
\eeq
{}From (\ref{phi_M}), we can read off non-perturbative corrections to
(\ref{phi_ele}) and (\ref{phi_mag}). At the weak coupling limit
$\tau\ra i\infty$ ($h(\tau)\ra 1$), (\ref{phi_M}) becomes
\beq
\prod_{a=1}^{N_c}(v+\phi_a)=v^{N_c},
\eeq
which agrees with the semi-classical result (\ref{phi_ele}).
On the other hand, in the strong coupling limit $\tau\ra 0$
($h(\tau)\ra -1$), (\ref{phi_M}) becomes
\beq
\prod_{a=1}^{N_c}(v+\phi_a)=(v+2m_S)^{N_f-N_c}
\prod^{2N_c}_{j=N_f+1}(v-m_j+2m_S),
\eeq
which is consistent with the semi-classical result in the magnetic theory
(\ref{phi_mag}).

\subsection{Duality in $\N=1$ MQCD}
In order to obtain $\N=1$ MQCD, we rotate one of the NS5-branes in the
$\N=2$ brane configuration to $w=x^8+x^9i$
direction. Here we only consider the baryonic branch on which
the curve is factorized into the two components of (\ref{fact}):
\beq
&C_R:&t=\alpha(v),\\
&C_L:&t=\beta(v).
\eeq
The functions
 $\alpha(v)$ and $\beta(v)$ are given in (\ref{alpha}) and (\ref{beta}).
The rotated curve is given as in Ref.~\citen{HOO}:
\beq
&C_R:&t=\alpha(v),~ w=\mu \hat v+\rho.,
\label{rotate_e}\\
&C_L:&t=\beta(v),~ w=0.
\label{rotate_e2}
\eeq
As argued in Ref.~\citen{HOO}, the meson VEV 
can be obtained from the value of $w$,
setting $t=0$ in (\ref{rotate_e}).
Eliminating $\hat v$ in (\ref{rotate_e}), we obtain
\beq
t=(1+h)\left(\frac{1}{\mu}(w-\rho)\right)^{N_c},
\label{t_ele}
\eeq
which means that the meson VEV is
\beq
M=\diag (\underbrace{\rho,\cdots,\rho}_{N_c},
0,\cdots,0).
\eeq
This result agrees with the field theory result (\ref{meson_e}).
Note that this is not a good variable
in the strong coupling limit $h\ra -1$,
since the right hand side of (\ref{t_ele}) vanishes.

Then, how can we obtain the meson VEV in the magnetic theory?
As explained in the last subsection, we should read it
from the same Riemann surface as above.
Actually there is another candidate for the meson VEV.

Let $\hat w\equiv w-\mu \hat v-\rho$, and rewrite the curve as
\beq
&C_R:&t=\alpha(v),~ \hat w=0,
\label{rotate_m1} \\
&C_L:&t=\beta(v),~ \hat w=-\mu \hat v-\rho.
\label{rotate_m}
\eeq
We propose that the meson VEV in the magnetic theory can be read from 
the value of $\hat w$ setting $t=0$ in (\ref{rotate_m}).
Eliminating $\hat v$ in (\ref{rotate_m}), we obtain
\beq
t=(1-h)\left(\frac{-1}{\mu}(\hat w+\rho)\right)^{N_f-N_c}
\prod^{2N_c}_{j=N_f+1}\left(\frac{-1}{\mu}(\hat w+\rho)-m_j\right),
\eeq
from which we interpret the meson VEV in the magnetic theory as
\beq
N=\diag (0,\cdots,0,\underbrace{-\rho,\cdots,-\rho}_{N_f-N_c},
\underbrace{-\sigma_{N_f+1},\cdots,-\sigma_{2N_c}}_{2N_c-N_f}),
\eeq
where $\sigma_i\equiv \rho+\mu m_i$. This form is exactly what we have
expected in (\ref{meson_m}).

\subsection{Toward Duality in Field Theory}
What we have shown in the last subsection is that the meson VEV
 for both the electric and magnetic theories can be read from the
 identical Riemann surface.
Here we want to try to explain how to interpret this result in the
field theory.

We proposed in \S \ref{N=1SQCD} that
the electric and magnetic theories have the same matter content
with couplings chosen as follows.
\beq
\begin{array}{ccc}
\hline\hline
 &\mbox{electric} &\mbox{magnetic}\\
\hline
\mbox{gauge coupling} & \tau& -1/\tau\\
\mbox{quark mass}& m_i & m'_i \\
\mbox{adjoint mass}&\mu&-\mu\\
\hline
\end{array}
\nonumber
\eeq
Notice that 
we should always consider the weak coupling
region whenever we compare the M-theory description
with the field theory description.

First we set $\tau\sim i\infty$ and
consider the brane configuration for the weak coupling electric
theory.
 As in (\ref{rotate_e}), we rotate $C_R$ to satisfy
the asymptotic condition $w\sim \mu v$. (Fig. \ref{fig1})

\begin{center}
\unitlength=.4mm
\begin{picture}(100,80)(0,0)
\epsfxsize=4cm
\put(0,-10){\epsfbox{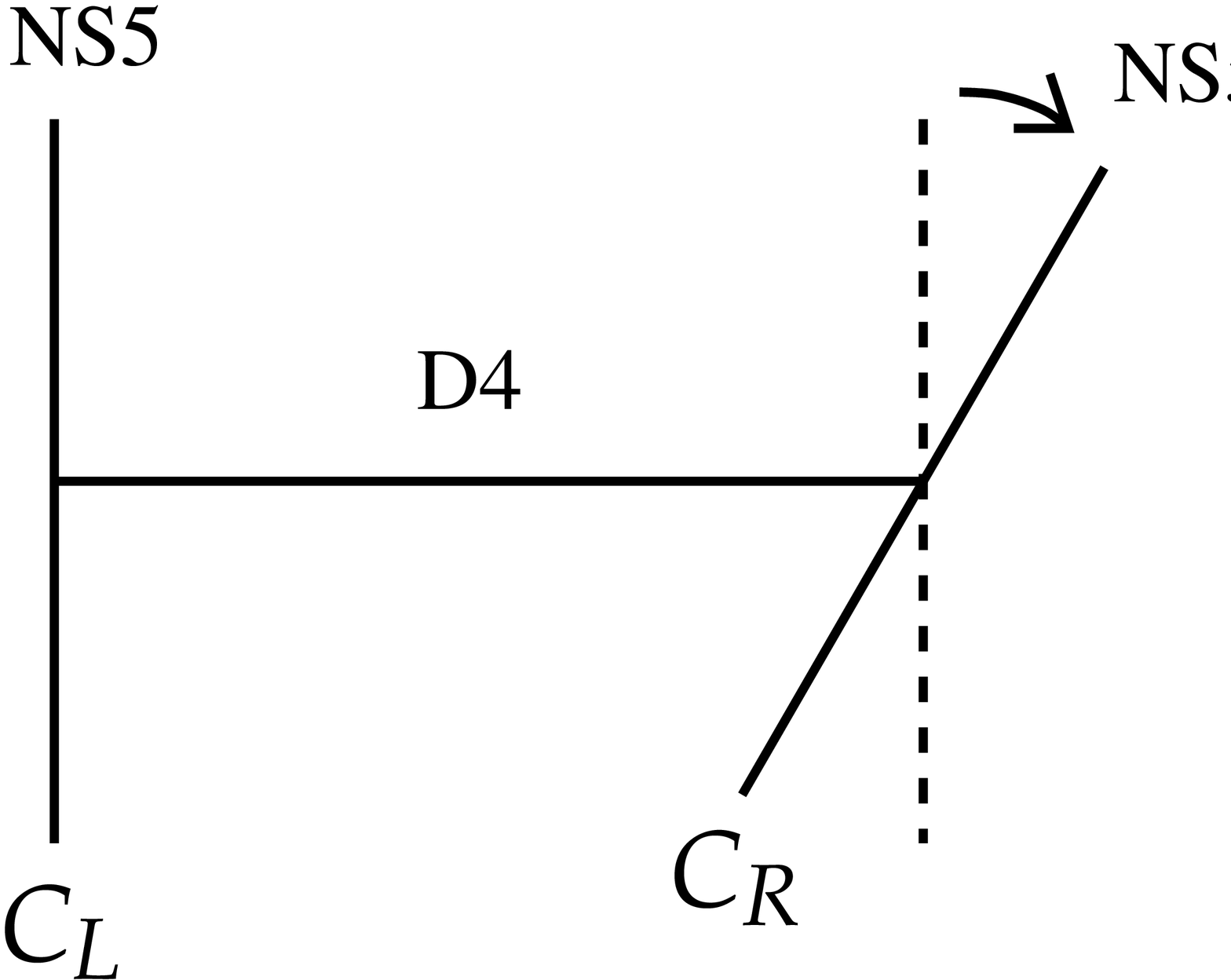}}
\end{picture}
\end{center}
\vspace{2ex}
\begin{center}
\refstepcounter{figure}
\label{fig1}
{Fig. \thefigure. } the weak coupling electric theory
\end{center}
\vspace{2ex}

On the other hand, if we wish to obtain the brane configuration
for the proposed magnetic theory in the same way as above,
we should move $\tau$
to $-1/\tau=\tau'\sim i\infty$, i.e. exchange the two NS5-branes, and then
rotate $C_L$ with the asymptotic condition $w\sim -\mu v$.
(Fig. \ref{fig2})
\begin{center}
\unitlength=.4mm
\begin{picture}(100,80)(0,0)
\epsfxsize=3.8cm
\put(0,-10){\epsfbox{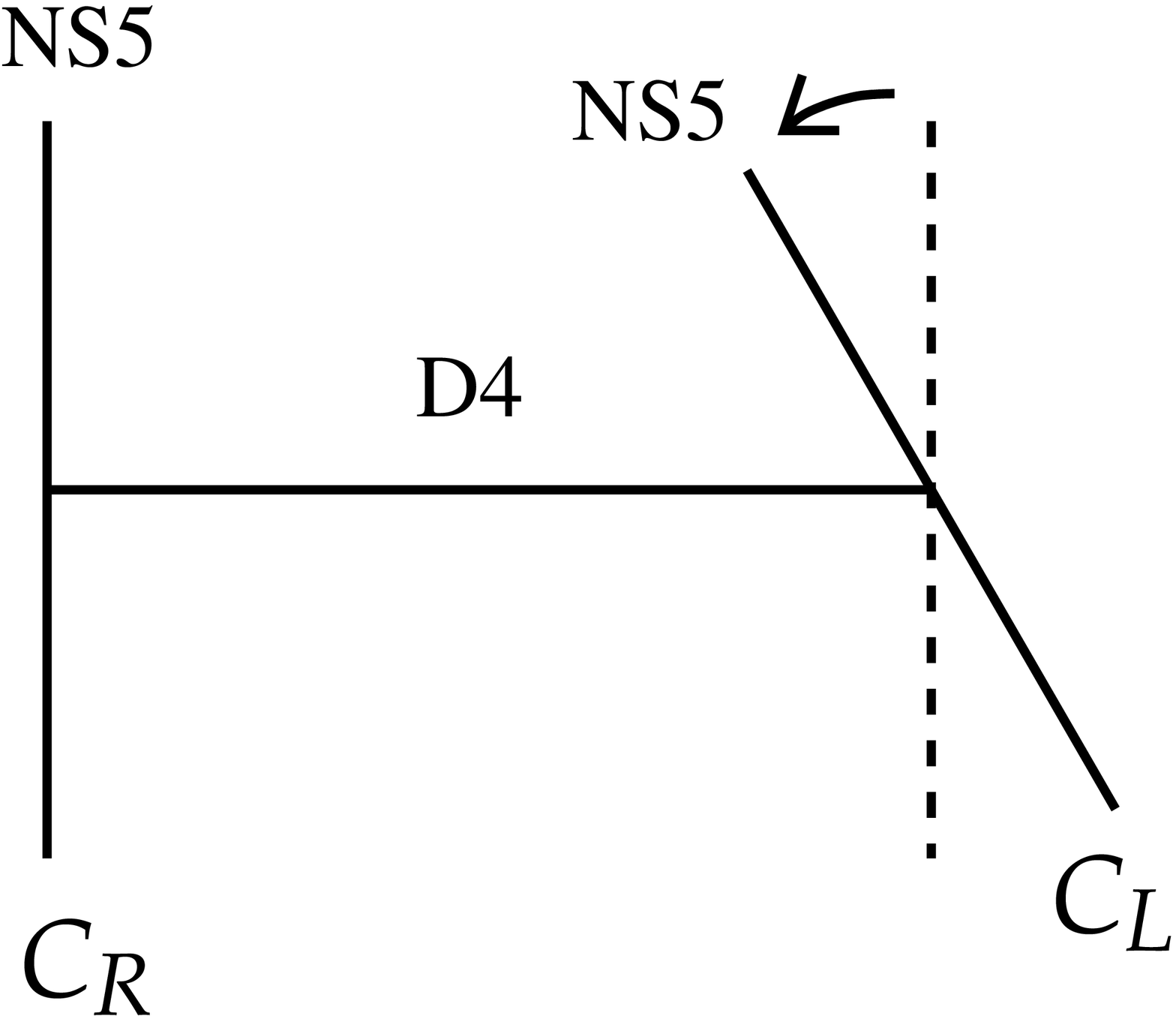}}
\end{picture}
\end{center}
\vspace{2ex}
\begin{center}
\refstepcounter{figure}
{Fig. \thefigure. } the weak coupling magnetic theory
\label{fig2}
\end{center}
\vspace{2ex}

In order to compare this theory with the weak coupling
electric theory, we restore the value
of $\tau$ to $\tau\sim i\infty$. As a result, we obtain the brane
configuration for the magnetic theory proposed in the field theory
approach, in which $C_L$ is rotated in the opposite direction
as $C_R$ in Fig. \ref{fig1}. (Fig. \ref{fig3})

\begin{center}
\unitlength=.4mm
\begin{picture}(100,80)(0,0)
\epsfxsize=4cm
\put(0,-10){\epsfbox{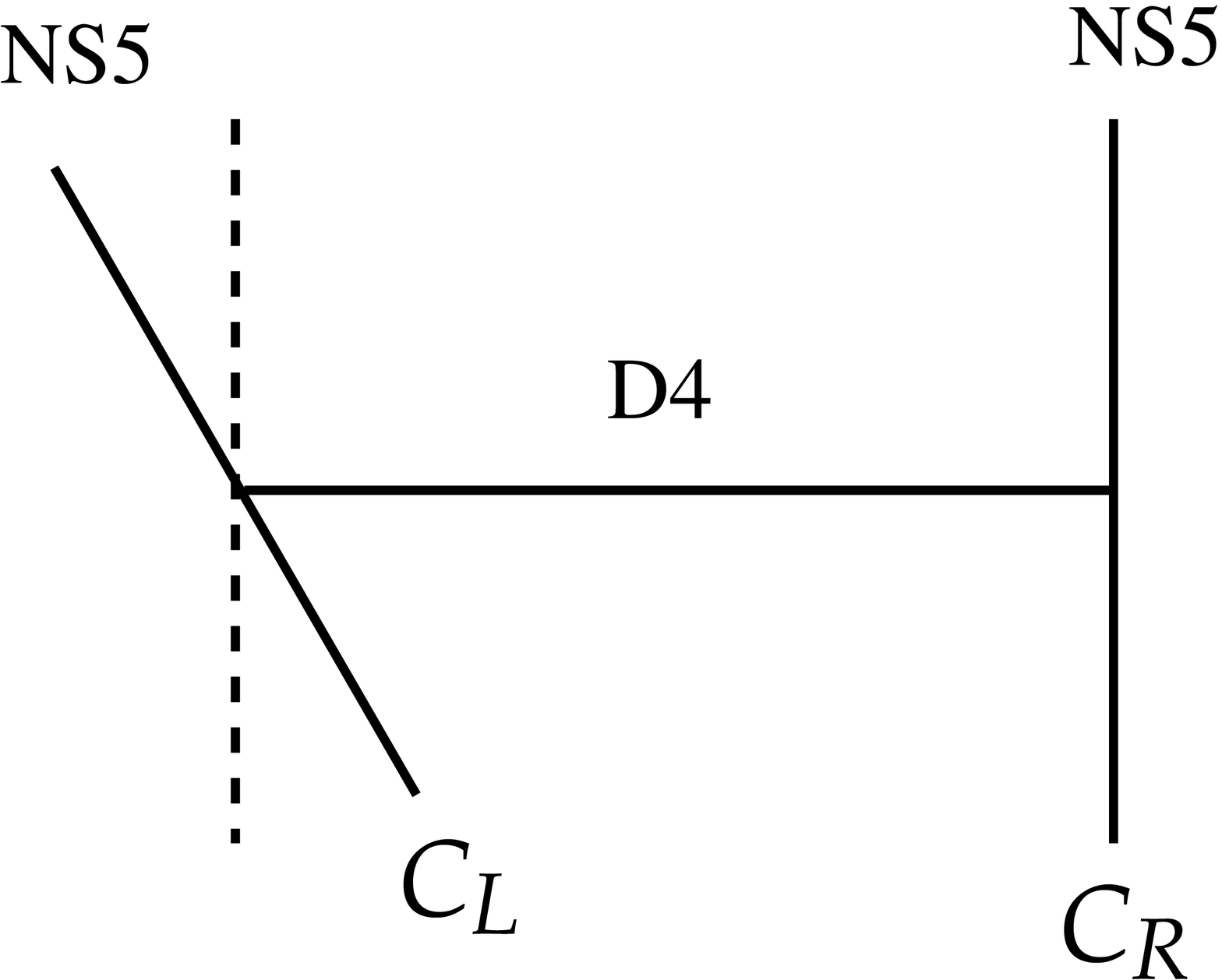}}
\end{picture}
\end{center}
\vspace{2ex}
\begin{center}
\refstepcounter{figure}
\label{fig3}
{Fig. \thefigure. } the strong coupling magnetic theory
\end{center}
\vspace{2ex}

This brane configuration is nothing but that given in
(\ref{rotate_m1}) and (\ref{rotate_m}), replacing $\hat w$ with $w$.
 Now it has become clear the reason why
we have interpreted the meson VEV in the magnetic theory
as the value of $\hat w$ instead of $w$ in the last subsection,
in order to reproduce the field theory results.
However, this brane configuration is not the same as that in 
the electric theory, and this theory is {\it not} equivalent to the
electric theory. If we set $\hat w=x^8+x^9 i$ instead of setting
$w=x^8+x^9 i$, the K\"ahler potential and the higher derivative terms
 in MQCD would be changed,  
since they pick up the background space-time metric,
 which is sensitive to the coordinate change
$\hat w\ra w$.\cite{dBHOO,LW,W1} 
The two configurations in Fig. \ref{fig1}
 and Fig. \ref{fig3} are actually the same
as the complex manifolds. So only the holomorphic structures, such as
VEVs for the holomorphic gauge invariant operators, are expected to
be correspondent.
If we wish to obtain an exact duality, we should again rotate
the two NS5-branes in Fig. \ref{fig3}
 to obtain the same brane configuration as in
the electric theory  Fig. \ref{fig1}.
 Therefore the magnetic theory which is exactly
equivalent to the electric theory will have complicated
non-holomorphic terms in its Lagrangian.

Now consider taking the limit $\mu\ra\infty$. In this limit, $\hat w$
is no longer a good coordinate to parameterize the $w=x^8+x^9 i$
direction, and therefore the meson VEV $N$ in the magnetic theory
loses its physical meaning. Hence we should introduce a new variable
instead of $N$ to parameterize the vacuum moduli space
in the magnetic description. 
This suggests that we must include the gauge singlet meson field $M$,
which is needed in Seiberg's duality,
in the massless matter content of the magnetic theory.

Let us make some comments on the coupling flow
(see also Ref.~\citen{HMS}).
Seiberg's duality is believed to be true only in the low energy limit.
The reason is that both the electric and magnetic theories seem to be 
asymptotically free (for $3/2 N_c<N_f<3N_c$), so 
we will be able to observe the differences in the high energy region.
On the other hand, we have seen that the electric and magnetic theories
are exactly equivalent in MQCD. How can these situations be consistent?
The resolution is as follows.
It is believed that there is an IR fixed point
in the gauge coupling flow.\cite{S}
So if we tune the bare coupling to be stronger than the IR fixed point,
the theory becomes asymptotically non-free.
In our situation, both the electric and magnetic theories are
regularized by the finite $\N=2$ theory at an energy scale higher than
$\mu$ and $m_i$.
The gauge coupling is fixed to be the bare value $\tau$
at this high energy region.
As we have discussed in (\ref{mag}),
the bare couplings for the electric and magnetic theories are related
as $\tau=-1/\tau'$. This fact suggests that
if the electric gauge coupling is weaker than the IR fixed point,
then the magnetic gauge coupling is stronger than the IR fixed point.
Therefore if the electric theory is an asymptotically free theory,
the magnetic theory should be asymptotically non-free.
When we move to the high energy region, where
the electric theory is weak coupling, the magnetic coupling
becomes stronger and stronger, and so
there is no apparent discrepancy in Seiberg's duality even in the
high energy region.

\section{BPS states and duality}
\label{sect4}
In the last section, we defined the electric and magnetic theories
as an effective world volume theory of an M5-brane wrapped on the
same Riemann surface. In this formulation, duality is indeed manifest,
but the interpretation in the usual field theory language
 is not so clear. Hence, in order to understand more detailed 
structures of duality, it is important to
examine the particle content of the theory and
determine the correspondence of the particles
under duality.

In this section, we return to the analysis in $\N=2$ $SU(N_c)$
MQCD with $2N_c$ flavors,
and examine BPS states in the theory.
We will see that there are W-bosons (massive vector multiplets)
 and quarks (hypermultiplets) in the weak coupling
electric theory, as expected,
and wo find that the elementary particles in the magnetic theory
appear as magnetic monopoles in the electric theory.
Moreover, we find many exotic states which cannot be obtained
in the perturbative field theory.
We will also discuss the realizations of BPS states via
geodesics in the Riemann surface
$\Sigma$, and apply the technique to the $N_c=2$ case.

In these analyses, M-theoretical viewpoints are essential,
and it is quite non-trivial to interpret the results 
in the IIA picture. As an example, we will consider an interesting
phenomenon in type IIA string theory in \S \ref{sect4.5}.

\subsection{Multi Taub-NUT space}

In this section we embed the Riemann surface $\Sigma$, on which
the M5-brane is wrapped, into multi-Taub-NUT
space $Q$ \cite{W}.
$Q$ is a hyper-K\"ahler manifold, which possesses
complex structures $I,J$ and $K$ satisfying
\beq
&&I^2=J^2=K^2=-1,\\
&&IJ=-JI=K,~JK=-KJ=I,~KI=-IK=J.
\eeq
As a complex manifold with respect to the complex structure $I$,
 $Q$ can be written as
\beq
YZ=\prod^{2N_c}_{j=1}(\hat v-m_j)
\equiv\prod^{2N_c}_{j=1}\left(v-m_j+(1-h(\tau))m_S\right),
\label{YZ}
\eeq
where $Y,Z$ and $v$ are complex variables which can be related to
the real coordinate $(x^4,x^5,x^6,x^{10})$ as \cite{Hit,NOYY}
\beq
Y&=&e^{-(x^6+x^{10}i)}\prod_{j=1}^{2N_c}
   \sqrt{\,|\vec{x}-\vec{x_j}|-(x^6-x_j^6)\,},\\
Z&=&e^{x^6+x^{10}i}\prod_{j=1}^{2N_c}
   \left(\sqrt{\,|\vec{x}-\vec{x_j}|+(x^6-x_j^6)\,}\,
   \left(\frac{\hat v-m_j}{|\hat v-m_j|}\right)\right),\\
\hat v&=&x^4+x^5i,
\eeq
where $\vec{x}=(x^4,x^5,x^6),~ m_j=x^4_j+x^5_j i$.
$\vec{x_j}=(x^4_j,x^5_j,x^6_j)$ is the position of 
NUT singularities, which can be interpreted as the position of
D6-branes in the type IIA picture.
In these coordinates, the metric on $Q$ is given as \cite{Haw}
\beq
ds^2=V d\vec{x}^2+\frac{1}{V}(dx^{10}+\vec{\omega}\cdot d\vec{x})^2,
\eeq
where
\beq
&&V=1+\sum\frac{1}{2|\vec{x}-\vec{x_i}|}\\
&&\vec{\nabla}\times\vec{\omega}=\vec{\nabla}V.
\eeq
The K\"ahler form $K$ and the holomorphic 2-form $\Omega$ are
\cite{Hit,NOYY}
\beq
K&=&i\,V d\hat v\wedge d\ol{\hat v}+
\frac{i}{V}\left(\frac{dY}{Y}-\delta\, d\hat v \right)
\wedge\overline{\left(\frac{dY}{Y}-\delta\, d\hat v \right)},
\label{kahler}\\
&&\delta\equiv\frac{1}{2}\sum_{j=1}^{2N_c}
\frac{\,x^6-x^6_j+|\vec{x}-\vec{x_j}|\,}{|\vec{x}-\vec{x_j}|\,(\hat v-m_j)},\\
\Omega&=&2\,d\hat v\wedge\frac{dY}{Y}.
\label{omega}
\eeq

We take $\Sigma$ to be
\beq
\sqrt{1-h(\tau)^2}\;(Y+Z)=2\prod^{N_c}_{a=1}(v+\phi_a),
\label{Sigma}
\eeq
which is equivalent, as a Riemann surface, to the curve in (\ref{ele}).

For later use, let us express $Q$ as a complex manifold with 
respect to another complex structure which is orthogonal to $I$.
The holomorphic variables with respect to the complex structure
$e^{I\theta}J$ are
\beq
\wt Y&=&e^{-(\wt x^4+(x^{10}-\gamma)\,i)}\prod_{j=1}^{2N_c}
   \sqrt{\,|\vec{x}-\vec{x_j}|-(\wt x^4-\wt x_j^4)\,},
\label{wtY}\\
\wt Z&=&e^{\wt x^4+(x^{10}-\gamma)\,i}\prod_{j=1}^{2N_c}
   \left(\sqrt{\,|\vec{x}-\vec{x_j}|+(\wt x^4-\wt x_j^4)\,}\,
   \left(\frac{\wt v-\wt m_j}{|\wt v-\wt m_j|}\right)\right),
\label{wtZ}\\
\wt v&=&-x^6+\wt x^5i, 
\label{wtv}
\eeq
where we have defined 
\beq
\wt x^4+\wt x^5i&=&e^{-i\theta}(x^4+x^5i), 
\label{wtx4x5}\\
\gamma&=&\sum_{j=1}^{2N_c}
 \arg \left(\wt v-\wt m_j+|\vec{x}-\vec{x_j}|-(\wt x^4-\wt x^4_j)
 \right),\\
\wt m_j&=&-x^6_j+\wt x^5_ji.
\eeq
Using these variables, $Q$ can be written as
\beq
\wt Y\wt Z=\prod_{j=1}^{2N_c}(\wt v-\wt m_j).
\label{TN}
\eeq
The K\"ahler form $\wt K$ and the holomorphic 2-form $\wt \Omega$ are
\beq
\wt K&=&i\,V d\wt v\wedge d\ol{\wt v}+
\frac{i}{V}\left(\frac{d\wt Y}{\wt Y}-\wt\delta\, d\wt v \right)
\wedge\overline{\left(\frac{d\wt Y}{\wt Y}-\wt\delta\, d\wt v \right)},
\\
&&\wt\delta\equiv\frac{1}{2}\sum_{j=1}^{2N_c}
\frac{\,\wt x^4-\wt x^4_j+|\vec{x}-\vec{x_j}|\,}{|\vec{x}-\vec{x_j}|\,(\wt
  v-\wt m_j)},\\
\wt\Omega&=&2\,d\wt v\wedge\frac{d\wt Y}{\wt Y}.
\eeq
It is straightforward to check the following relations:
\beq
\wt K&=&\Im\,(e^{-i\theta}\Omega),\\
{\rm Re}~\wt\Omega&=&{\rm Re}~(e^{-i\theta}\Omega),\\
\Im \wt \Omega &=& -K.
\eeq

\subsection{BPS states in MQCD}
\label{BPS}
BPS states in MQCD are realized as M2-branes ending on
the M5-brane.\cite{W} The M2-brane is decomposed as $\R\times\Sigma'$, where
$\R$ is the world line of the particle and 
$\Sigma'$ is a Riemann surface
holomorphically embedded in $Q$ with respect to the complex structure
$e^{I\theta}J$, and the boundary of $\Sigma'$ lies in
$\Sigma$.\cite{FS,HY,M}

The ele-mag charges for the BPS states can be read from the homology class
of the boundary $\partial\Sigma'$ in $\Sigma$.\cite{KLMVW,FS,MNS}
Let us explain this fact explicitly.

There is a self-dual 2-form field $B^+_2$, whose field strength
$H_3^+=dB^+_2$ is self-dual, on the M5-brane world volume.
When the M5-brane is wrapping a Riemann surface $\Sigma$,
$B^+_2$ should be expanded via harmonic forms on $\Sigma$
in order to pick up the massless modes in the four dimensional
effective theory.
If the genus of $\Sigma$ is $l$, we have $l$ holomorphic 1-forms
$\omega^J~(J=1,\cdots,l)$ on $\Sigma$.
The harmonic 1-forms are given by the real and imaginary parts of 
the holomorphic 1-forms.
Setting $\omega_{\ele}^J\equiv {\rm Re }~\omega^J$ and $\omega_{\mag}^J
\equiv \Im~\omega^J$, we have
\beq
B^+_2=\omega_{\ele}^J\wedge A_{\ele}^J
+\omega_{\mag}^J\wedge A_{\mag}^J,
\eeq
where $A_{\ele}^J$ and $A_{\mag}^J$ are 1-form
fields in $\R^4$.\footnote{
Note that there are no normalizable harmonic 0,2-forms, since the 
Riemann surface $\Sigma$ is not compact.}
Since the field strength of $B^+_2$ is self dual,
using the relation $\star\omega_{\ele}=\omega_{\mag}$, it follows that 
the field strengths of $A_{\ele}$ and $A_{\mag}$ are dual to each other:
$\star F_{\ele}= F_{\mag}$.

We choose a symplectic basis $\{\alpha_I,\beta_I\}$ of $H_1(\Sigma,\Z)$,
and we take $\omega^J$ to satisfy 
\beq
\int_{\alpha_I}\omega^J&=&\delta^{IJ}.
\eeq
Then, the matrix with entries
\beq
\int_{\beta_I}\omega^J&\equiv&\tau^{IJ}
\equiv \left(\frac{8\pi i}{g^2}\right)^{IJ}+\frac{\theta^{IJ}}{\pi},
\eeq
becomes the effective gauge coupling constant for the $U(1)^{l}$ theory,
 as in the Seiberg-Witten theory.\cite{SW,Ver,W}

The boundary of M2-branes ($\simeq \R\times\partial\Sigma'$) are strings
on the M5-brane and couple to the 2-form field \cite{T} as
\beq
S_{\rm int}\sim\int_{\sR\times\partial\Sigma'}B_2^+ .
\eeq
If the homology class of $\partial\Sigma'$ is
$n_e^I\alpha_I+n_m^I\beta_I\in H_1(\Sigma,\Z)$,
 putting all these data together,
we obtain
\beq
S_{\rm int}\sim
\left(n_e^J+n_m^I\frac{\theta^{IJ}}{\pi}\right)\int_{\sR} A_{\ele}^J
+ n_m^I\left(\frac{8\pi}{g^2}\right)^{IJ}\int_{\sR} A_{\mag}^J.
\eeq
{}From this, we can interpret $n_e+n_m\frac{\theta}{\pi}$ 
as the electric charges
and $n_m$ as the magnetic charges of the BPS state. 
Thus we have obtained the standard charge assignment in
Seiberg-Witten theory.\cite{SW}
Note that when $\theta\ne 0$, the electric charges are shifted from $n_e$ by
$n_m\frac{\theta}{\pi}$. This phenomenon is well known in the field theory
as the `Witten effect'.\cite{Weff}

\subsection{Construction of BPS states}
\label{QandW}

In this subsection, we first consider the weak coupling limit,
and find $\Sigma'$ for the
W-bosons and quarks in the multi-Taub-NUT space $Q$.
We also show that there are no other particles in this limit.
These result shows that
the theory is really $SU(N_c)$ gauge theory with $2N_c$ flavors
in the weak coupling region.\footnote{
We ignore Kaluza-Klein modes in our analysis.}
 Then we turn on the gauge coupling
and find more examples of BPS states.

Let us first consider the 
weak coupling limit $\tau\ra i\infty$, $h(\tau)\ra 1$.
In this limit, the M5-brane (\ref{Sigma}) reduces to
\beq
0=\prod^{N_c}_{a=1}(v+\phi_a).
\eeq
Thus,  $x^4$ and $x^5$
 are fixed to be constant on each connected component of $\partial\Sigma'$.
As mentioned above,
$\Sigma'$ is a Riemann surface holomorphically embedded in $Q$ with respect
to the complex coordinates
 $\wt Y,\wt Z,\wt v$ in (\ref{wtY}) $\sim$ (\ref{wtv}).
Let $p$ be a projection map from $\Sigma'$ to the $\wt v$-plane
induced by the canonical projection $(\wt Y,\wt Z,\wt v)\ra\wt v$.
Then $p(\partial\Sigma'_0)$
lies on
a fixed line which is parallel to the real axis, since $\wt x^5$ takes
a fixed value. Here $\partial\Sigma'_0$ denotes a
connected component of $\partial\Sigma'$.
Now we restrict our discussion on BPS states with finite mass.
Since the mass is proportional to the area of $\Sigma'$,
the closure of $\Sigma'$ should be compact.
Let us suppose that $p$ is not a constant map.
Then, since $p$ is holomorphic, $p$ is an open map.
Using a standard technique in topology, we can show that
the boundary of $p(\Sigma')$ lies in $p(\partial\Sigma')$.
But then, since $p(\partial\Sigma')$
is a union of parallel lines,
the image of $p$ will inevitably extend infinitely in the
$x^6$ direction,
contradicting the compactness of the closure of $p(\Sigma')$.
Thus we conclude that $p$ must be a constant map.


We classify $\Sigma'$ in two cases :

\noindent (i) ~$\wt v=\wt m_j$ for some $j$

\begin{figure}
\begin{center}
\unitlength=.4mm
\begin{picture}(100,60)(0,0)
\epsfxsize=4cm
\put(0,-15){\epsfbox{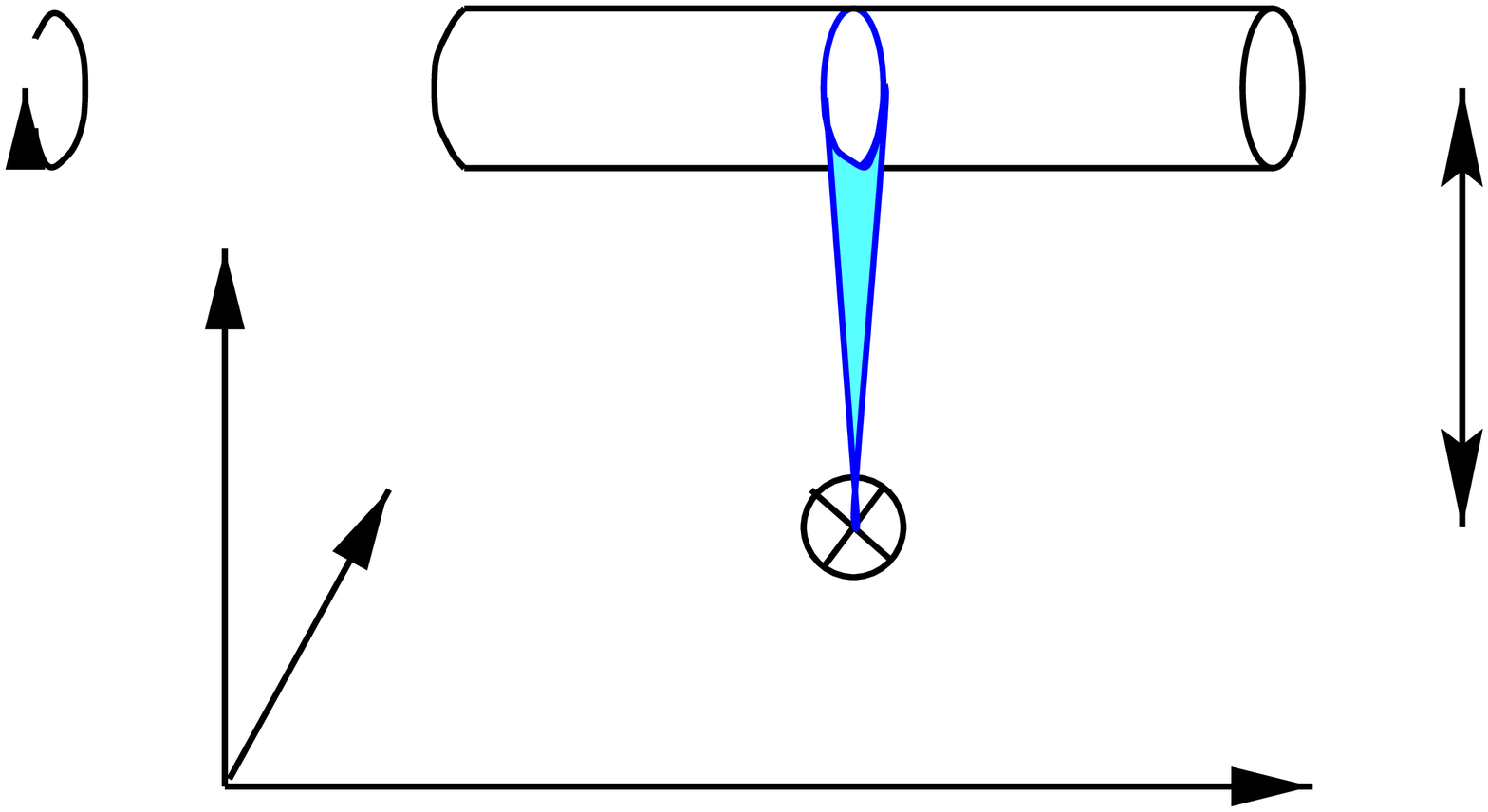}}
\put(-10,37){\makebox(0,0){$x^{10}$}}
\put(97,-13){\makebox(0,0){$x^{6}$}}
\put(32,12){\makebox(0,0){$\wt x^{5}$}}
\put(18,30){\makebox(0,0){$\wt x^{4}$}}
\put(133,20){\makebox(0,0){$\sim |\,\phi_a+m_j\,|$}}
\end{picture}
\end{center}
\vspace{6ex}
\begin{center}
\parbox{10cm}{
\caption{
Here $\otimes$ represents
the position of a NUT
singularity, and the cylinder stretching in the $x^6$ direction
represents a part of the M5-brane $\Sigma$. The disk which is 
caught by a NUT and winding around a cycle in $\Sigma$
is the M2-brane $\Sigma'$ for a quark.
}
\label{fig4}
}
\end{center}
\end{figure}

In this case, (\ref{TN}) implies $\wt Y=0$ or $\wt Z=0$.
{}From (\ref{wtY}) and (\ref{wtZ}), $\wt Y=0$ implies $\wt x^4>\wt x^4_i$
and $\wt Z=0$ implies $\wt x^4<\wt x^4_i$.
We can tune $\theta$ in (\ref{wtx4x5}), in order that $\Sigma'$ intersects
$\Sigma$. In this case, $\Sigma'$ is a disk and
the BPS state in four-dimensional effective theory
makeup a hypermultiplet.\cite{HY,M} As a result, we have obtained
$2N_c$ flavors of quarks with mass $\propto |\,\phi_a+m_j\,|$.
(Fig. \ref{fig4})

\vspace{2ex}

\noindent (ii)~ $\wt v=\mbox{constant}\ne\wt m_j$ for all $j$


In this case, (\ref{TN}) implies $\wt Y\ne 0$ and $\wt Z\ne 0$.
We must tune $\theta$ in (\ref{wtx4x5}) in order that $\Sigma'$
can intersect with two cycles in $\Sigma$.
In this case, $\Sigma'$ is a cylinder and the corresponding BPS state
constitutes a vector multiplet \cite{HY,M}. As a result, we have obtained
W-bosons with mass $\propto |\,\phi_a-\phi_b\,|$.
When we take $\forall\phi_a=0$, they will become massless
and form an $SU(N_c)$ gauge multiplet together with the $U(1)^{N_c-1}$
fields given in \S \ref{BPS}. (Fig. \ref{fig5})

\begin{center}
\unitlength=.4mm
\begin{picture}(100,60)(0,0)
\epsfxsize=4cm
\put(0,-15){\epsfbox{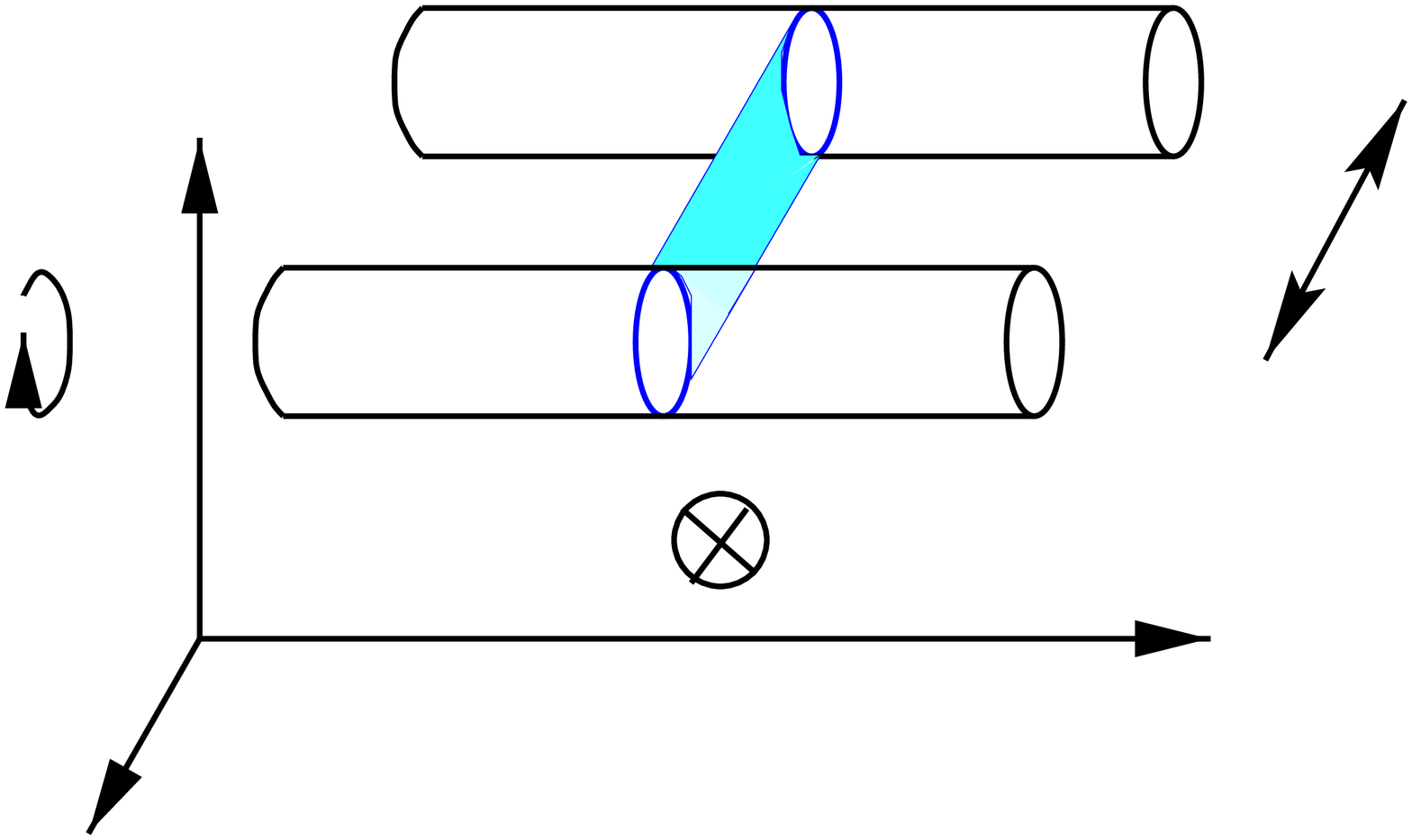}}
\put(12,42){\makebox(0,0){$\wt x^{5}$}}
\put(-10,23){\makebox(0,0){$x^{10}$}}
\put(93,0){\makebox(0,0){$x^{6}$}}
\put(0,-15){\makebox(0,0){$\wt x^{4}$}}
\put(125,27){\makebox(0,0){$\sim |\,\phi_a-\phi_b\,|$}}
\end{picture}
\end{center}
\vspace{5ex}
\begin{center}
\refstepcounter{figure}
{Fig. \thefigure. } W-boson
\label{fig5}
\end{center}
\vspace{2ex}

Next we switch on the gauge coupling.
It is not easy to find $\Sigma'$ explicitly in a generic brane configuration.
Thus we only give
several examples with a specific choice of $\phi_a$
and $\vec{x_j}$.

We use the following notation:
\beq
C(\hat v)&\equiv&
\prod^{N_c}_{a=1}(v+\phi_a)\equiv \hat v^{N_c}+s_1 \hat v^{N_c-1}
+\cdots+s_{N_c},\\
B(\hat v)&\equiv&
\prod^{2N_c}_{j=1}(\hat v-m_j)\equiv \hat v^{2N_c}+w_1
\hat v^{2N_c-1}+\cdots+w_{2N_c},\\
\wt B(\hat v)&\equiv&\sqrt{\,(1-h^2)\,B(\hat v)}.
\eeq
We consider the case in which $\forall x_j^6=0$, $-1<h(\tau)<1$
and $\forall s_a, \forall w_j\in\R$.
We take the complex structure with $\theta=0$
for the M2-brane and consider $\Sigma'$ satisfying $\wt v=0$, that
is, $x^5=x^6=0$.
Then the intersection of $\Sigma$ and $\Sigma'$ is given as
\beq
&&x^5=x^6=0,\\
&&C(x^4)=\wt B(x^4)\,\cos\,x^{10}.
\label{dyon}
\eeq

We can find various states by constructing
 graphs of $C(x^4)$ and $\pm\wt
B(x^4)$.
For example, typical graphs for dyons, W-bosons and
quarks are given as in Fig. \ref{fig6}.

\hspace*{4ex}
\parbox{4.5cm}{
\unitlength=.4mm
\begin{picture}(80,80)
\epsfxsize=3.8cm
\put(0,0){\epsfbox{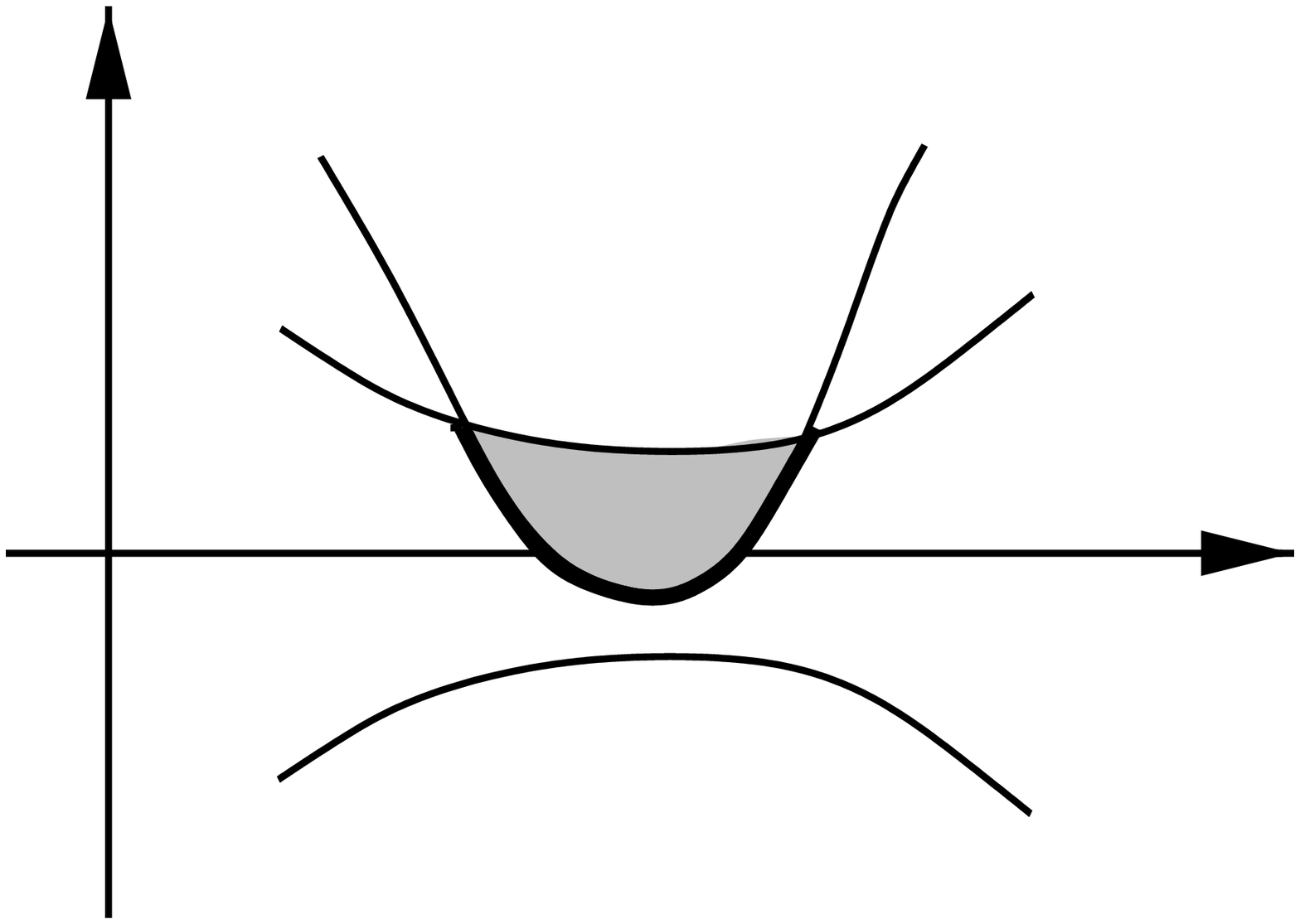}}
\put(83,5){\makebox(0,0){$-\wt B$}}
\put(85,50){\makebox(0,0){$\wt B$}}
\put(75,64){\makebox(0,0){$C$}}
\put(98,23){\makebox(0,0){$x^4$}}
\put(50,0){\makebox(0,0){dyon}}
\end{picture}
}
\parbox{4.5cm}{
\unitlength=.4mm
\begin{picture}(80,80)
\epsfxsize=3.8cm
\put(0,0){\epsfbox{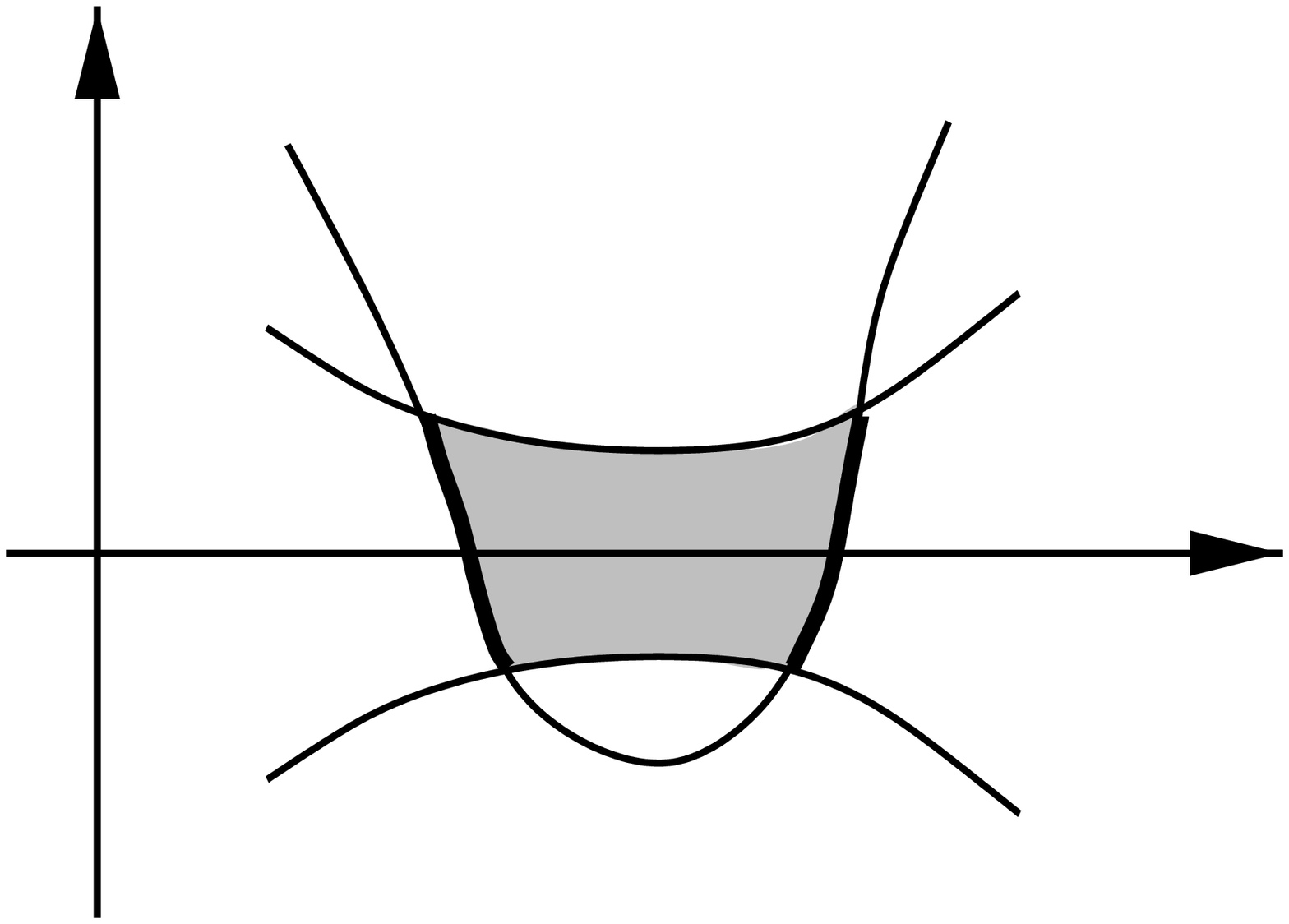}}
\put(50,0){\makebox(0,0){W-boson}}
\end{picture}
}
\parbox{4.5cm}{
\unitlength=.4mm
\begin{picture}(80,80)
\epsfxsize=3.8cm
\put(0,0){\epsfbox{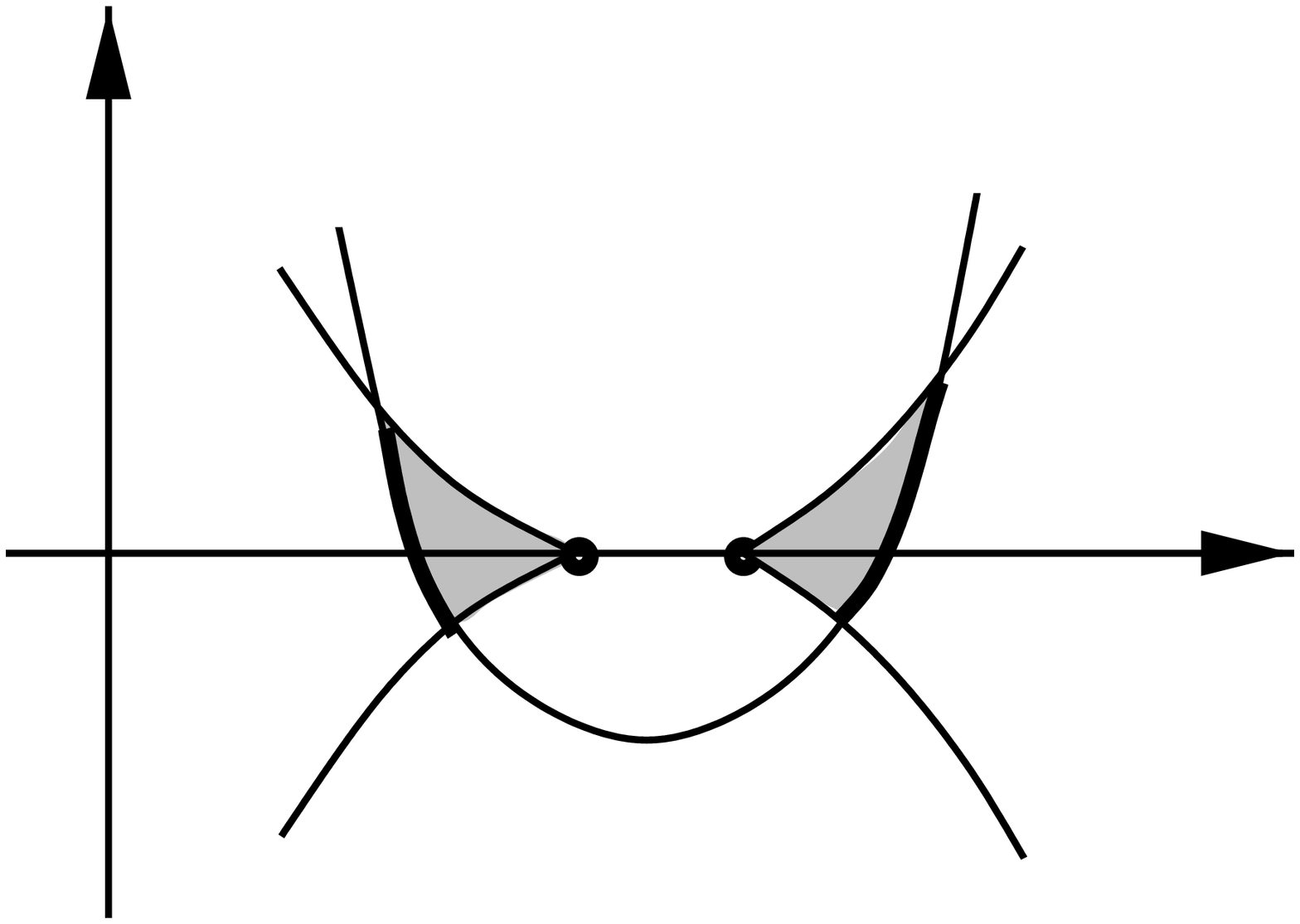}}
\put(50,0){\makebox(0,0){quarks}}
\end{picture}
}
\vspace{2ex}
\begin{center}
\refstepcounter{figure}
\label{fig6}
{Fig. \thefigure.}
\end{center}
\vspace{2ex}
\noindent
Note that solutions of $C(x^4)=\pm\wt B(x^4)$
are branch points on the real axis of the $\hat v$-plane, and
zero points of $\wt B$ are the NUT singularities.
We can see from these figures that when we tune the moduli parameters
$\phi_a$
or the mass parameters $m_j$ to obtain massless states,
two branch points must meet, and the Riemann surface $\Sigma$
becomes singular.
These figures also allow for the visualization of
decay patterns of the particles by moving
$\phi_a$ and $m_j$. For example, the W-boson in Fig. \ref{fig6}
will decay into two
quarks, as in Fig. \ref{fig6},
 tuning $m_j$ such that the two NUTs meet on the $\wt v=0$ plane.
Similarly, tuning $\phi_a$, the W-boson in Fig. \ref{fig6}
will decay into the dyon in Fig. \ref{fig6}, and
some other particles which are required in order to be consistent
 with charge conservation.
We can never find all the particles to decay 
in this picture, since they are, as argued in Ref.~\citen{HY}, mutually
non-local.

There are much more complicated and physically interesting 
examples, as shown in Fig. \ref{fig7}.
\begin{center}
\parbox{4.5cm}{
\unitlength=.4mm
\begin{picture}(80,80)
\epsfxsize=4cm
\put(0,0){\epsfbox{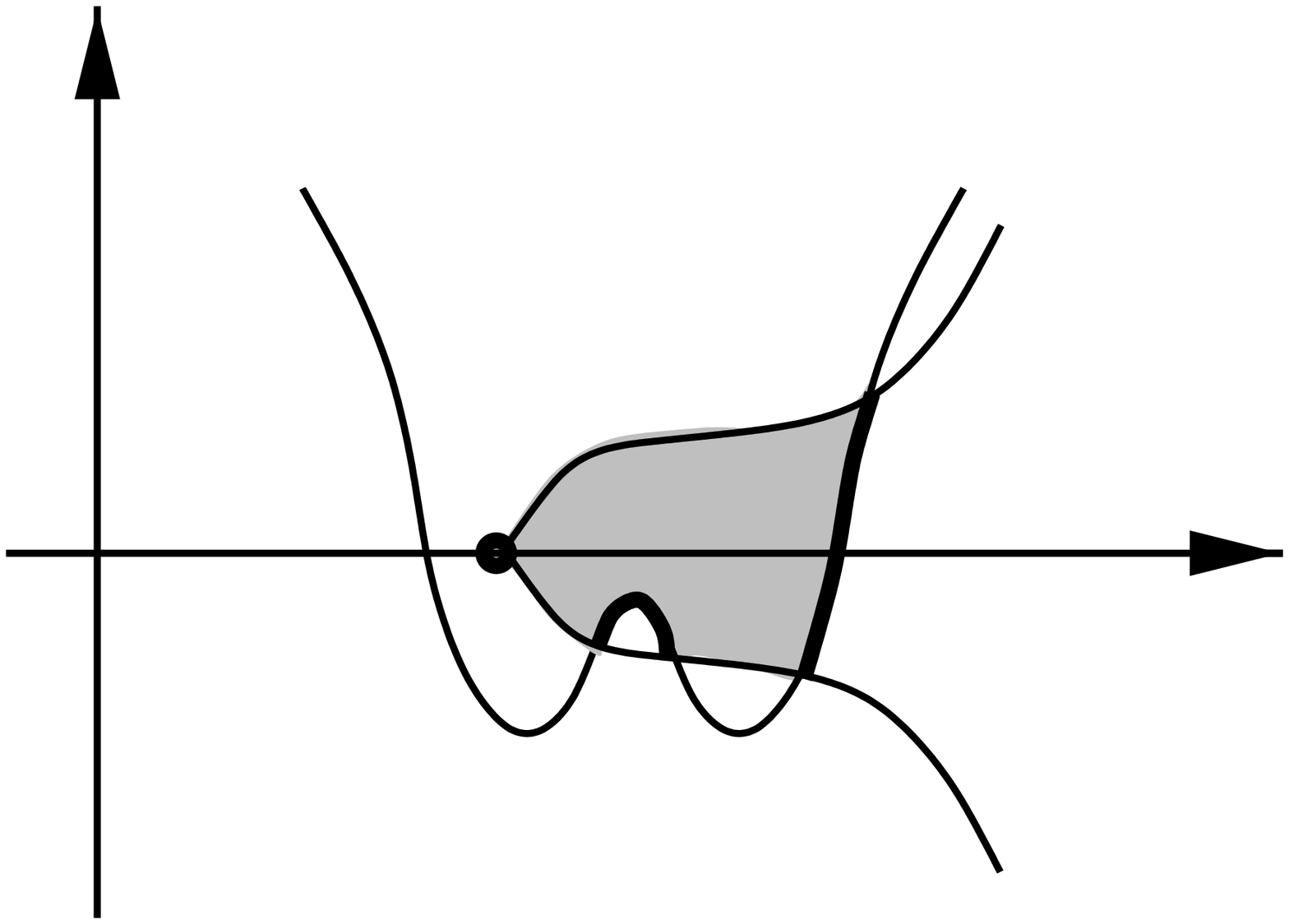}}
\put(60,0){\makebox(0,0){exotic W-boson ?}}
\end{picture}
}
\parbox{4.5cm}{
\unitlength=.4mm
\begin{picture}(80,80)
\epsfxsize=4cm
\put(0,0){\epsfbox{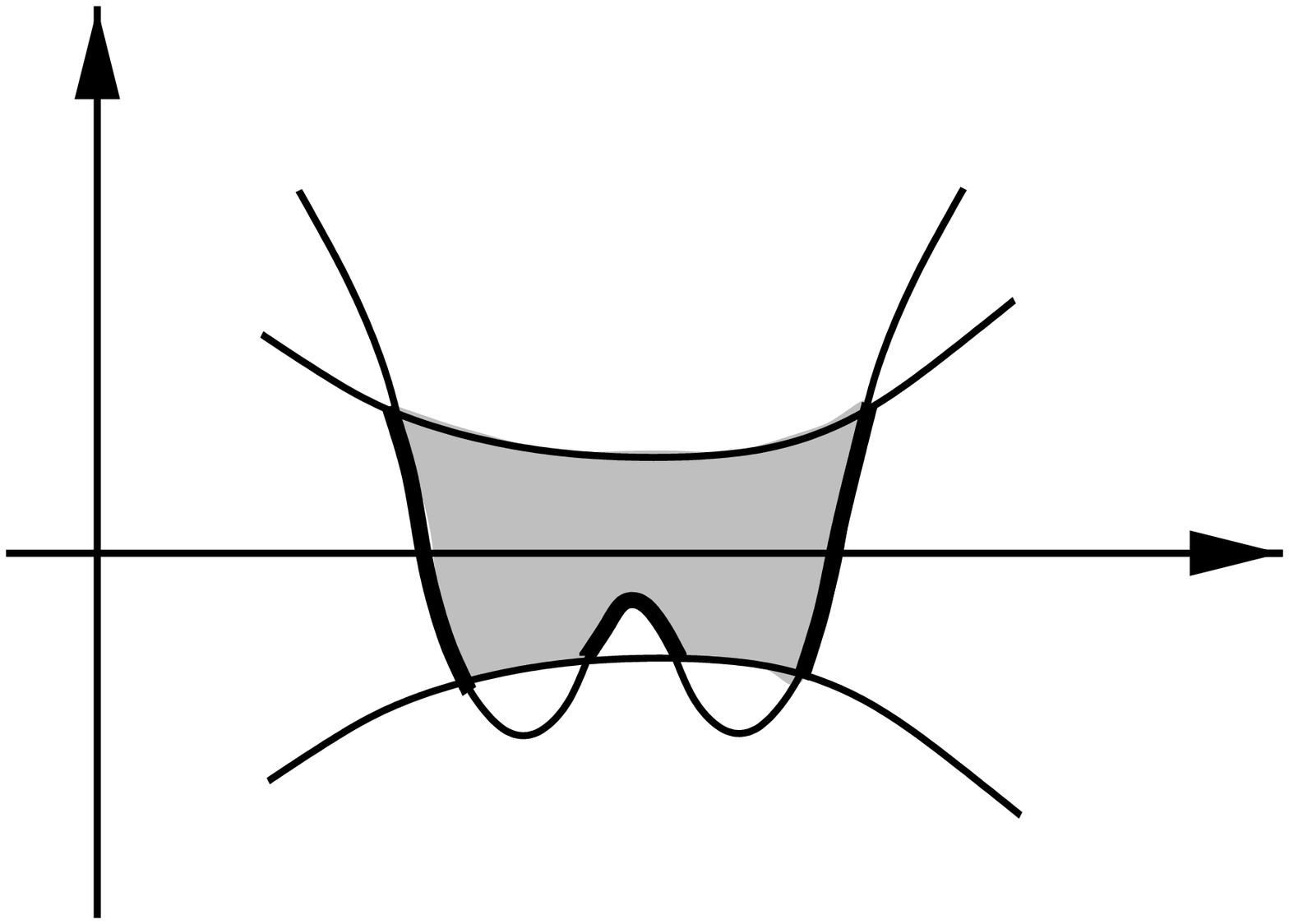}}
\put(65,0){\makebox(0,0){exotic particle ?}}
\end{picture}
}
\end{center}
\vspace{3ex}
\begin{center}
\refstepcounter{figure}
\label{fig7}
{Fig. \thefigure.}
\end{center}
\vspace{2ex}
\noindent
MQCD predicts the existence of these exotic particles in $\N=2$ SQCD.
There are examples of $\Sigma'$ with more than two holes, which
may correspond to multiplets with spin $3/2$ or higher.
They carry non-parallel electric and magnetic charges
and thus there seem to be no classical solutions corresponding to 
these particles in the $\N=2$ field theory.\cite{AY,FH} 

Although it would be quite interesting to explore further,
we wish to return to the original subject
in order to be faithful to the title. (See Ref.~\citen{HHS} for a
recent progress in $\N=4$ SYM.)

\subsection{Magnetic theory in electric theory}

In the usual field theory description, one defines a theory giving
a Lagrangian. We refer to fields in the Lagrangian as elementary
fields, and refer to
 the corresponding particles as elementary particles.
The elementary particles
in the electric theory constitute an $SU(N_c)$ vector multiplet and
$2N_c$ flavors of hypermultiplets.
In the last subsection, we have found those particles
at the weak coupling limit in the electric description of MQCD.
If we consider the strong coupling limit, we find
the elementary particles in the magnetic theory, since the theory is
weak coupling in the magnetic description.
It is obvious that the elementary particles in the electric theory
and the magnetic theory are not identical.
So, if one chooses the elementary particles in the electric
theory to write down the Lagrangian,
we should interpret the elementary particles in the magnetic theory
as solitons in the field theory.

What is remarkable in MQCD is that the elementary particles and solitons
are treated on an equal footing.
Both the electric and magnetic particles appear as M2-branes ending on
the M5-brane.
Thanks to this feature,
we can treat the electric and magnetic theories
simultaneously,
and it becomes possible to
analyze the particle correspondence under duality.

Now let us analyze the correspondence of the ele-mag charges
and make sure that the elementary particles in the magnetic theory
 are indeed magnetic monopoles in the electric description.
For simplicity, we set all the quark mass parameters to zero
and take $\phi_a=\phi\,e^{\frac{2\pi i}{N_c}a}$.
 Then the Seiberg-Witten
curve (\ref{ele}) becomes
\beq
t^2-2(v^{N_c}+\phi^{N_c})\,t+(1-h^2)\,v^{2N_c}=0.
\eeq
The branch points in the $v$-plane are given by
\beq
v_{a\pm}=\phi_a\left(\frac{-1\pm\sqrt{1-h^2}}{h^2}\right)^{\frac{1}{N_c}}.
{}~~~~~~(a\in \Z~(\mod N_c))
\label{branch}
\eeq

\begin{figure}
\unitlength=.4mm
\begin{center}
\begin{picture}(150,150)
\epsfxsize=6cm
\put(0,0){\epsfbox{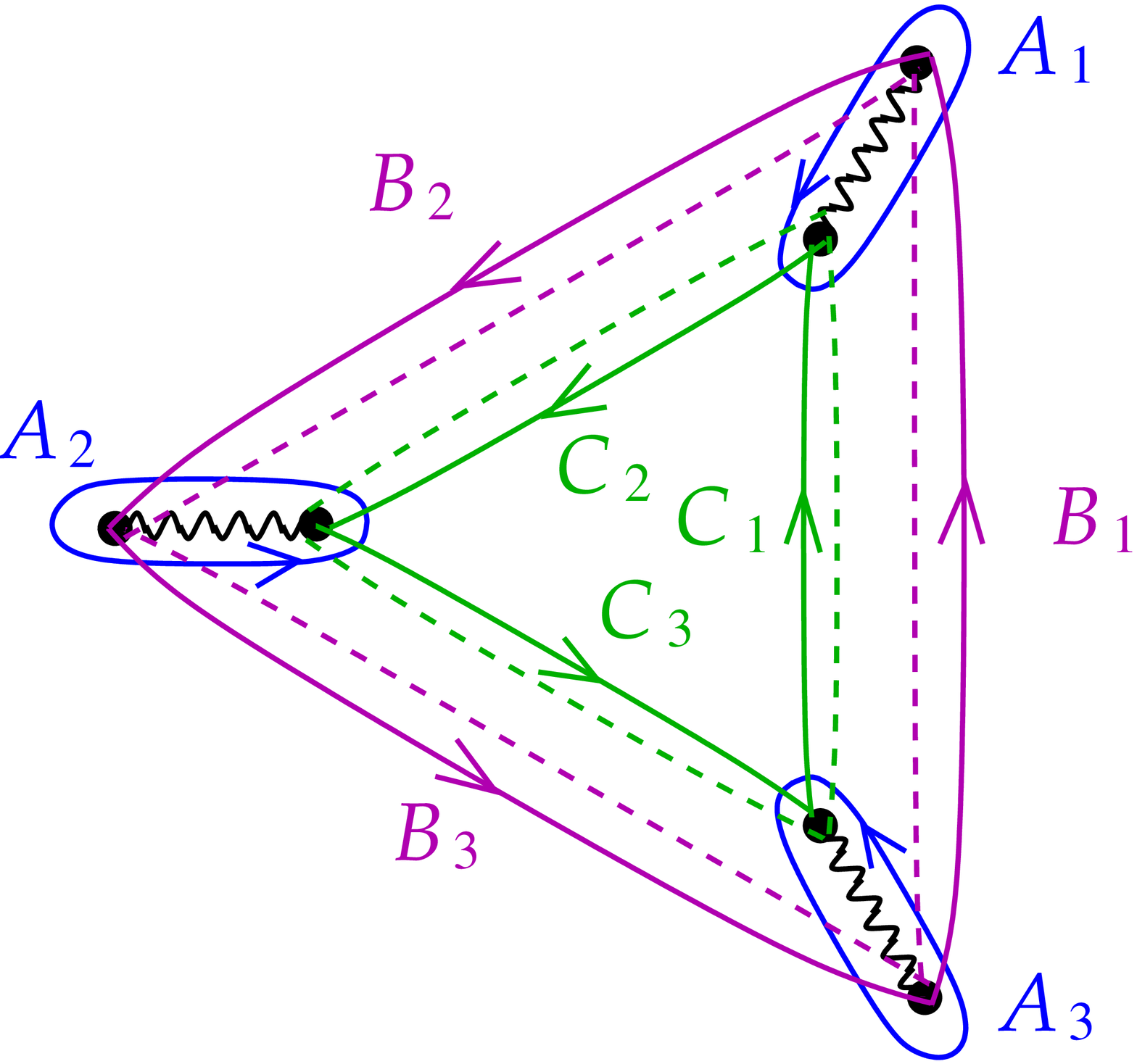}}
\end{picture}
\end{center}
\vspace{2ex}
\begin{center}
\parbox{10cm}{
\caption{
The $A_a,B_a,C_a$-cycles for the case
$N_c=3$ and $0<h<1$ in $v$-plane. The wavy lines
are branch cuts stretched between the branch points
$v_{a+}$ and $v_{a-}$.
}
\label{fig8}
}
\end{center}
\end{figure}

We define the $A_a$-cycle, $B_a$-cycle and $C_a$-cycle
 to be cycles which 
encircle ($v_{a+}$, $v_{a-}$), ($v_{(a-1)+}$, $v_{a+}$),
and ($v_{(a-1)-}$, $v_{a-}$) in the $v$-plane, respectively,
 as shown in
Fig. \ref{fig8}.

\vspace{1ex}

The intersection numbers for these cycles are given as
\beq
\begin{array}{ccrccr}
A_a\cdot B_a&=&1,~&A_{a-1}\cdot B_a&=&-1,\\
A_a\cdot C_a&=&1,~&A_{a-1}\cdot C_a&=&-1,\\
B_{a-1}\cdot B_a&=&-1,~&C_{a-1}\cdot C_a&=&1,
\end{array}
\eeq
and the others are all zero.
Of course, they are not all independent, satisfying the following
relations:
\beq
&&\sum_{a=1}^{N_c}A_a=\sum_{a=1}^{N_c}B_a=\sum_{a=1}^{N_c}C_a=0,\\
&&A_{a-1}+A_a=B_a-C_a.
\eeq
We choose the symplectic basis $\{\alpha_I,\beta_I|I=1,\cdots,N_c-1\}$
of $H_1(\Sigma,\Z)$ as
\beq
\alpha_I&\equiv&A_I,\\
\beta_I&\equiv&\sum_{a=1}^{I}(B_a-A_{a-1}),
\eeq
where $A_0\equiv A_{N_c}$. It is easy to check that they satisfy
\beq
&&\alpha_I\cdot\alpha_J=\beta_I\cdot\beta_J=0,\\
&&\alpha_I\cdot\beta_J=\delta_{IJ}.
\eeq

At the weak coupling limit in the electric description ($h\ra 1$),
the branch points $v_{a+}$ and
$v_{a-}$ approach each other, and $\alpha$-cycles
become the vanishing cycle. The quarks and W-bosons considered in the
last subsection are constructed with M2-branes, whose boundary is
homotopic to the linear combinations of the $\alpha$-cycles in
$\Sigma$ and thus is electrically charged 
in the convention given in \S \ref{BPS}. 

The elementary particles in the magnetic theory are
constructed similarly in the strong coupling limit 
in the electric description ($h\ra -1$).
So let us move on to the strong coupling region; that is,
we exchange the two NS5-branes in the type IIA picture. 
We consider moving $h$ from
the weak coupling region ($h\sim 1$) 
to the strong coupling region ($h\sim -1$)
along the real axis. Since there is a singularity at $h=0$, which is
due to the collision of two NS5-branes, we should deform the path near
$h\sim 0$ to avoid the singularity.

It would be easier to consider this in the $h^2$-plane.
The above prescription is equivalent to moving $h^2$ once around $0$.
When $h^2$ is nearly $0$, the branch points are at
$v_{a+}\sim \phi_a (-1/2)^{\frac{1}{N_c}}$ and
$v_{a-}\sim \phi_a (-2/h^2)^{\frac{1}{N_c}}$. 
While moving $h^2$ around $0$, $v_{a-}$ moves to $v_{(a-1)-}$, and
the $v_{a+}$ are fixed.

\vspace{3ex}
\begin{center}
\parbox{4cm}{
\unitlength=.4mm
\begin{picture}(100,80)(0,0)
\epsfxsize=3cm
\put(10,0){\epsfbox{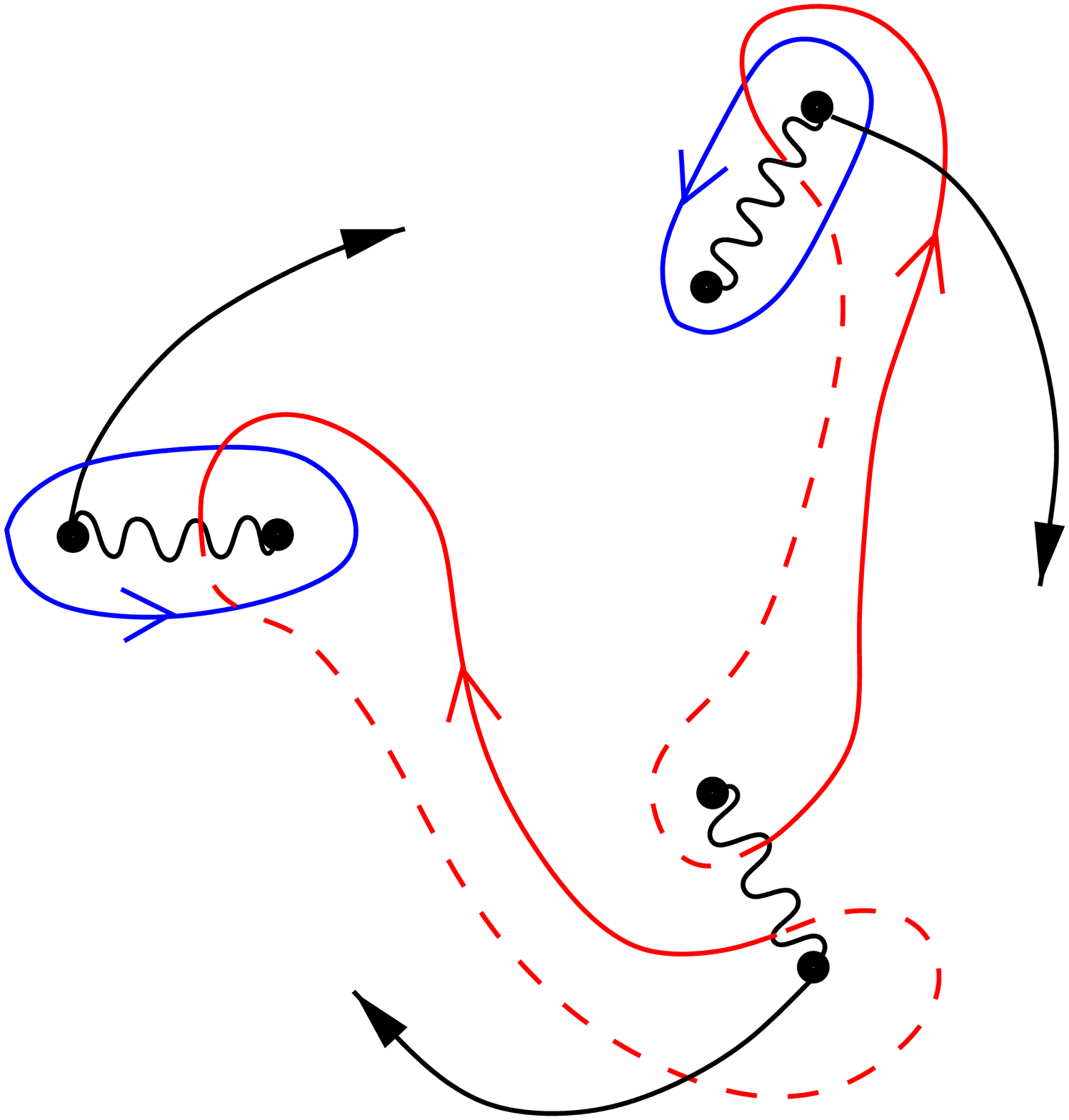}}
\end{picture}
}
$\Longrightarrow$
\parbox{4cm}{
\unitlength=.4mm
\begin{picture}(100,90)(0,0)
\epsfxsize=3cm
\put(10,0){\epsfbox{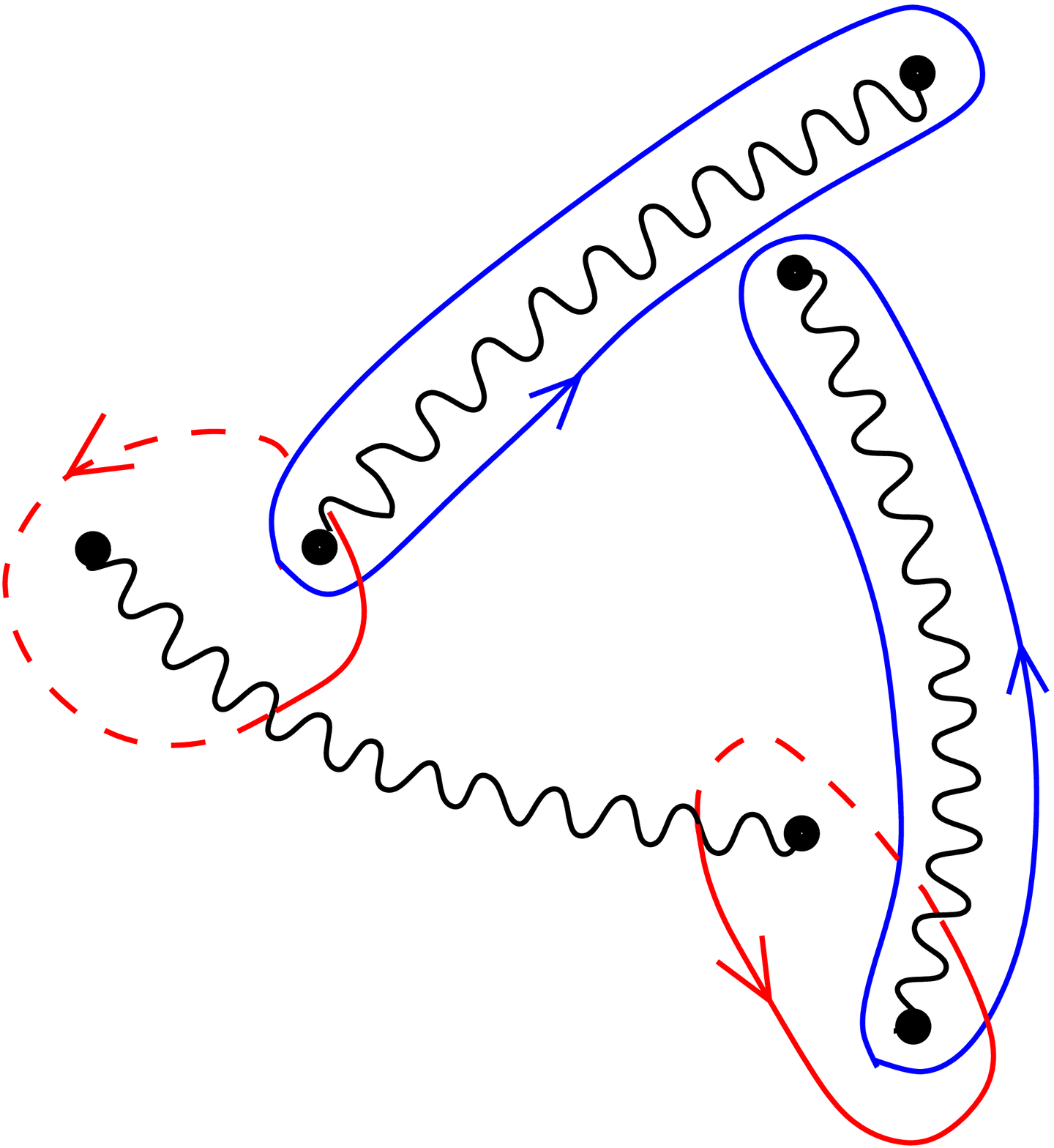}}
\end{picture}
}
\end{center}
\vspace{2ex}
\begin{figure}
\begin{center}
\parbox{10cm}{
\caption{
The motion of $\alpha$-cycles and $\beta$-cycles
via brane exchange. It can be easily seen that the role of $\alpha$-cycles
and $\beta$-cycles are exchanged.}
}
\end{center}
\end{figure}
\vspace{2ex}

We define the $A'_a$-cycle, $B'_a$-cycle and $C'_a$-cycle
as in Fig. \ref{fig10}.
These play the same role as $A_a,B_a,C_a$-cycle after
brane exchange.
\vspace{2ex}
\unitlength=.4mm
\begin{center}
\begin{picture}(150,150)
\epsfxsize=6cm
\put(0,0){\epsfbox{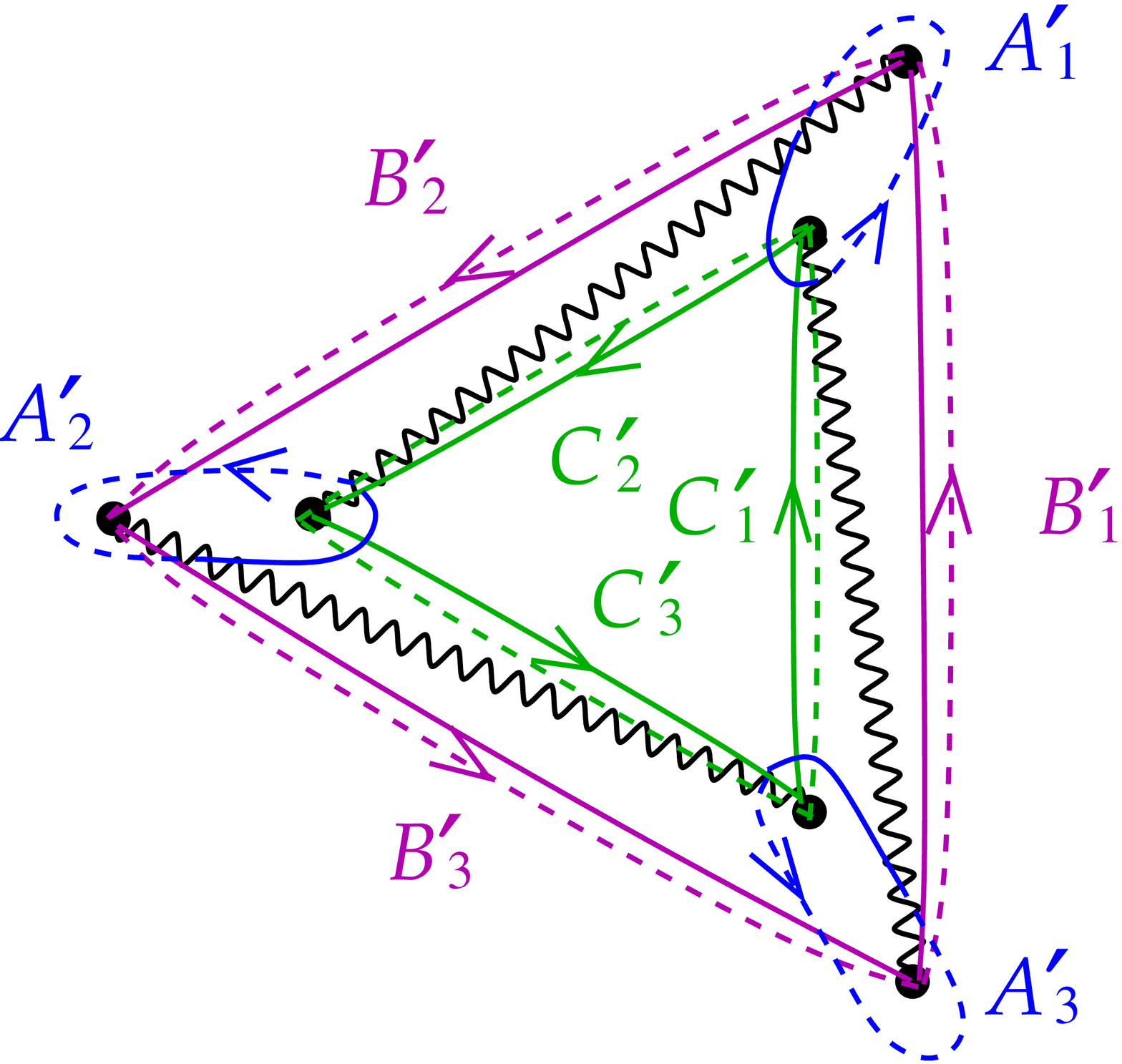}}
\end{picture}
\end{center}
\begin{center}
\refstepcounter{figure}
\label{fig10}
{Fig. \thefigure.}
\end{center}
\vspace{2ex}
The $A_a,B_a,C_a$-cycles are expressed as 
\beq
A_a&=& A'_a-B'_a,\\
B_a&=& -B'_{a-1},\\
C_a&=& C'_a.
\eeq

The symplectic basis which is canonical for the magnetic description
is
\beq
\alpha'_I&\equiv&A'_I,\\
\beta'_I&\equiv&\sum_{a=1}^{I}(B'_a-A'_{a-1}),
\eeq
where $I=1,\cdots,N_c-1$ and $A'_0\equiv A'_{N_c}$.
The relation between the two symplectic bases is
\beq
\alpha_I&=&\alpha'_I-\alpha'_{I-1}-\beta'_{I}+\beta'_{I-1},\\
\beta_I&=&\sum_{J=I}^{N_c-1} \alpha'_J,
\eeq
where $\beta'_0\equiv 0$, $\alpha'_0\equiv -\sum_{J=1}^{N_c-1}\alpha'_J$.
The charge assignments in these bases are related as
\beq
n_e^I&=& -\sum_{J=I}^{N_c-1}n_m^{'J},
\label{ne}\\
n_m^I&=& n_e^{'I}-n_e^{'I-1}+n_m^{'I}-n_m^{'I-1},
\label{nm}
\eeq
where we have defined
$n_e^{'0}\equiv 0$, $n_m^{'0}\equiv -\sum_{J=1}^{N_c-1}n_m^{'J}$.

The elementary particles in the magnetic description are
 dual quarks and dual W-bosons whose boundaries in the M2-branes are
$\partial\Sigma'= \pm A'_a$ and $\partial\Sigma'=  A'_a-A'_b$,
respectively. So they have $U(1)^{N_c-1}$ charges with
$n_e'\ne 0$ and $n_m'=0$. Equations (\ref{ne}) and (\ref{nm}) imply
that these particles are magnetic monopoles in the electric description.

As an example, let us consider the $N_c=3$ case.
Quarks and W-bosons in the electric theory have charges as follows:
\vspace{1ex}
\beq
\begin{array}{ccccc}
\hline
\hline
& n_e^1& n_e^2& n_m^1& n_m^2\\
\hline
& \pm 1&0&0&0\\ 
\mbox{quarks} & 0&\pm 1&0&0\\
& \pm 1&\pm 1&0&0\\
\hline
& \pm 1&\mp 1&0&0\\
\mbox{W-bosons}& \pm 1&\pm 2&0&0\\
& \pm 2&\pm 1&0&0\\
\hline
\end{array}
\eeq

\noindent
The dual quarks and dual W-bosons in the  magnetic theory can be
interpreted in the electric theory as magnetic monopoles with charges
given by (\ref{nm}):
\vspace{1ex}
\beq
\begin{array}{ccccc}
\hline
\hline
& n_e^1& n_e^2& n_m^1& n_m^2\\
\hline
\mbox{dual}& 0&0&\pm 1&0\\ 
\mbox{quarks} &0&0& 0&\mp 1\\
&0&0& \pm 1&\mp 1\\
\hline
\mbox{dual}&0&0& \pm 1&\pm 1\\
\mbox{W-bosons}& 0&0&\pm 1&\mp 2\\
&0&0& \pm 2&\mp 1\\
\hline
\end{array}
\eeq

\noindent
We can make a prediction for the field theory that
there exist these solitonic states in the electric theory
 at least in the strong coupling region.
These solitons will dominate in the strong coupling region and
form the dual $SU(N_c)$ multiplets.

Note, however, that we do not know whether these magnetic particles
also exist in the weak coupling region, since we may cross the
curves of marginal stability during the brane exchange,
and they may decay into some other particles \cite{SW,BF}. 
For the case $N_c=2$, there are no curves of marginal stability,
and the particle spectrum is predicted to be invariant under the
 duality group 
$SL(2,\Z)$.\cite{SW} But, as far as we know, it is still unclear
whether or not the particle spectrum is invariant under the duality group 
in the higher rank gauge theories, (see Ref.~\citen{AY} and \citen{CH}).
In \S \ref{geodesic}, we make a brief argument about these
points from the M-theoretical point of view.

\subsection{Brane conversion via brane exchange}
\label{sect4.5}
As a by-product of the analysis in the last subsection,
we observe an interesting phenomenon in type IIA string theory.

Recall that $\alpha$-cycles transform to $\beta$-cycles
via brane exchange. 
\vspace{5ex}
\begin{center}
\parbox{4cm}{
\unitlength=.4mm
\begin{center}
\begin{picture}(100,80)
\epsfxsize=3cm
\put(10,-5){\epsfbox{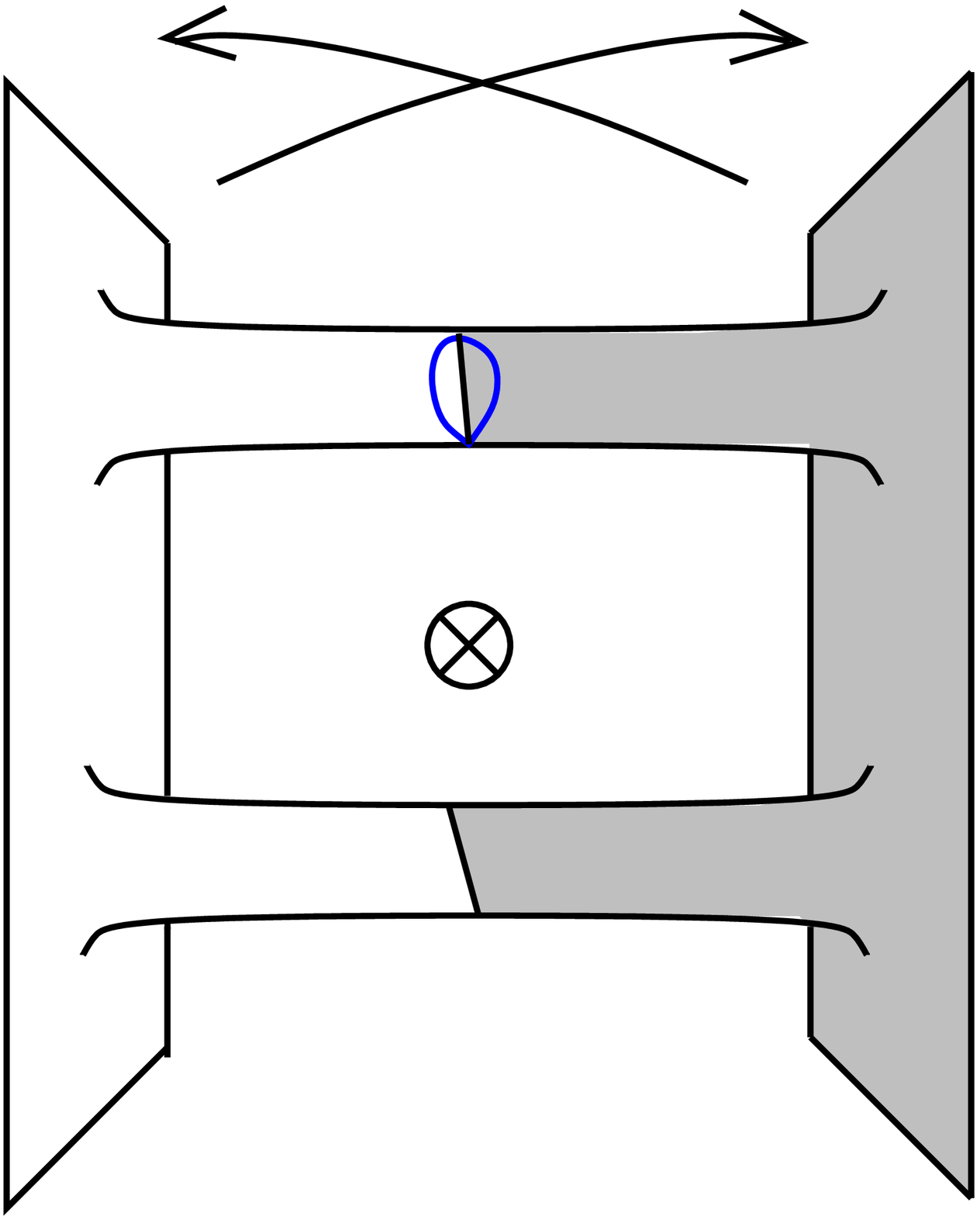}}
\end{picture}
\end{center}
}
$\Longrightarrow$
\parbox{4cm}{
\unitlength=.4mm
\begin{center}
\begin{picture}(100,80)
\epsfxsize=3cm
\put(10,-5){\epsfbox{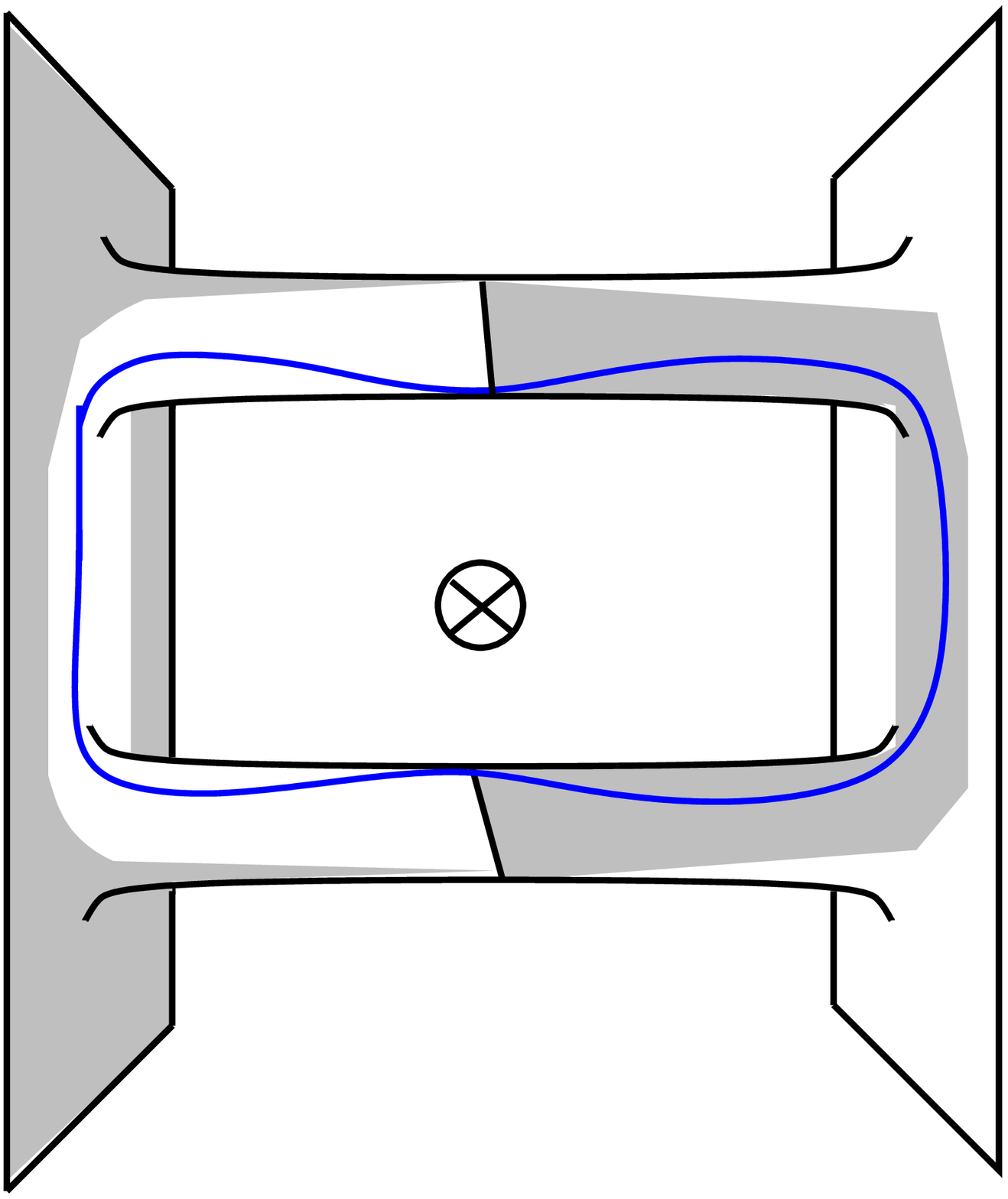}}
\end{picture}
\end{center}
}
\end{center}
\vspace{2ex}
\begin{figure}
\begin{center}
\parbox{10cm}{
\caption{
 M5-brane configurations 
for the $N_c=2$ case 
in 3-dimensional space $(x^4,x^5,x^6)$. Shadowed and unshadowed parts
represent two sheets in the $v$-plane.
Exchanging the brane,
 $\alpha$-cycles are transformed
into $\beta$-cycles.
}}
\end{center}
\end{figure}
\vspace{2ex}
Let us consider a quark in the weak coupling electric theory. 
As constructed in \S \ref{QandW}, the Riemann surface $\Sigma'$
for the quark is a disk caught by a NUT,
whose boundary winds around the $\alpha$-cycle of $\Sigma$
(Fig. \ref{fig4}).
It can be interpreted as a fundamental string stretched between
a D4-brane and a D6-brane in the IIA picture.
 Then, we move $\tau$ to the strong coupling
region and exchange the two NS5-branes. The quark is then interpreted
as a solitonic state in the weak coupling magnetic theory; that is,
the boundary of $\Sigma'$ winds around the $\beta$-cycle
of the brane exchanged Riemann surface $\Sigma$.
\vspace{5ex}
\begin{center}
\parbox{4cm}{
\unitlength=.4mm
\begin{center}
\begin{picture}(100,80)
\epsfxsize=3cm
\put(10,-5){\epsfbox{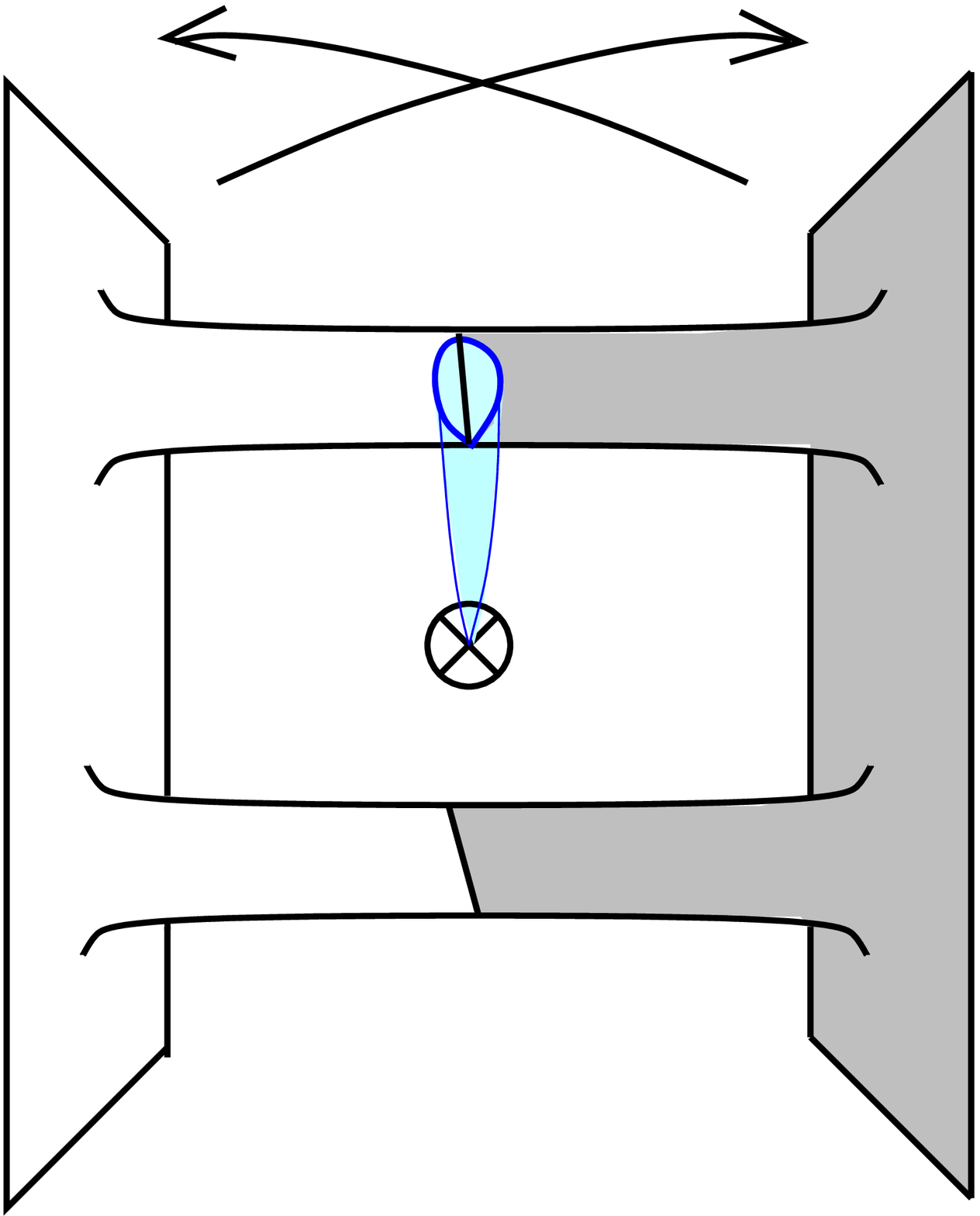}}
\end{picture}
\end{center}
}
$\Longrightarrow$
\parbox{4cm}{
\unitlength=.4mm
\begin{center}
\begin{picture}(100,80)
\epsfxsize=3cm
\put(10,-5){\epsfbox{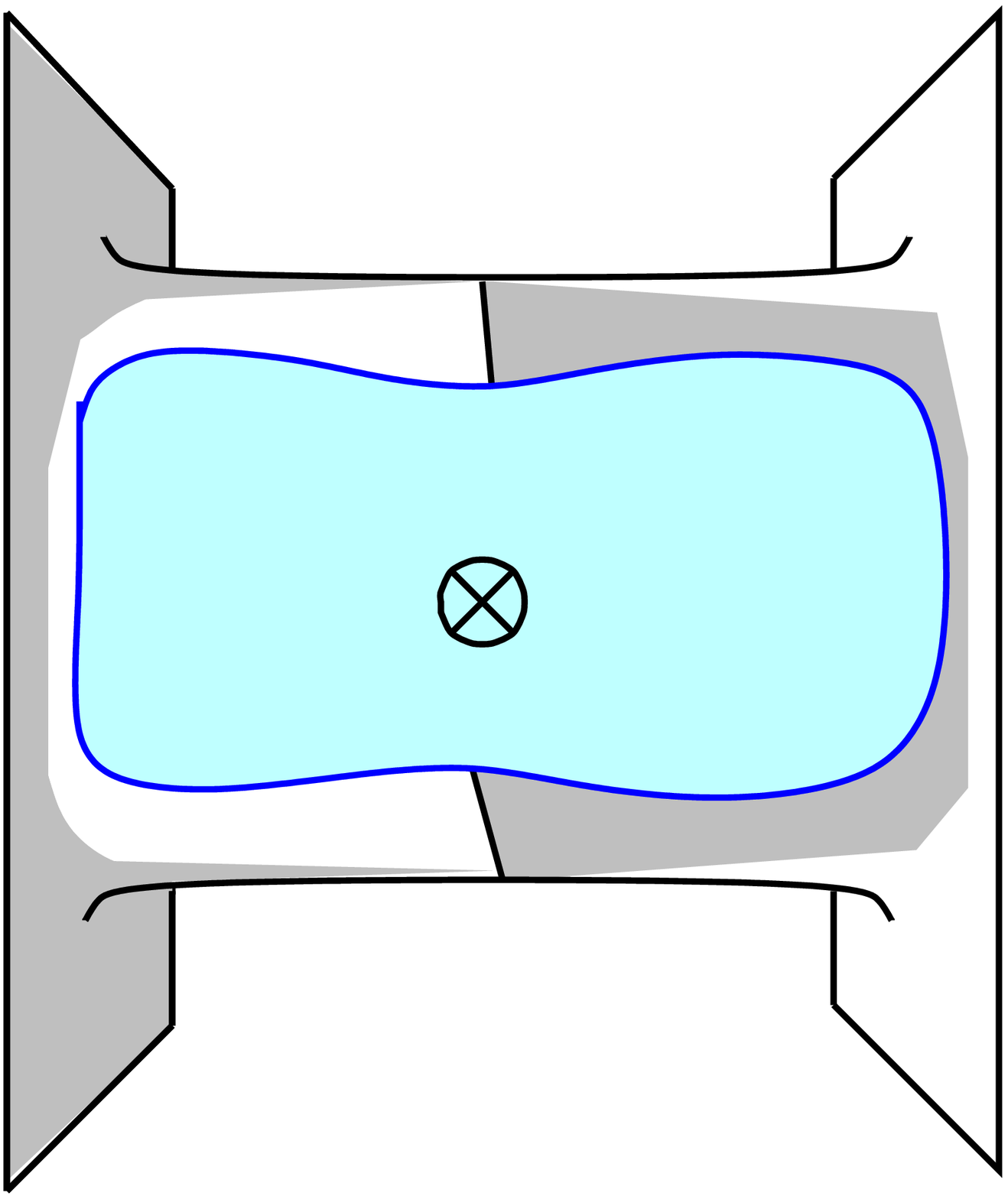}}
\end{picture}
\end{center}
}
\end{center}
\vspace{2ex}
\begin{center}
\refstepcounter{figure}
\label{fig12}
{Fig. \thefigure.}
\end{center}
\vspace{2ex}
Therefore the M2-brane for the quark
 in the strong coupling electric theory is
 interpreted as a D2-brane in the type IIA picture.
As a result, the fundamental string is converted into a D2-brane
via the exchange of two NS5-branes, as shown in Fig. \ref{fig13}.

\vspace*{3ex}
\begin{center}
\parbox{4cm}{
\unitlength=.4mm
\begin{center}
\begin{picture}(100,80)
\epsfxsize=3cm
\put(10,-5){\epsfbox{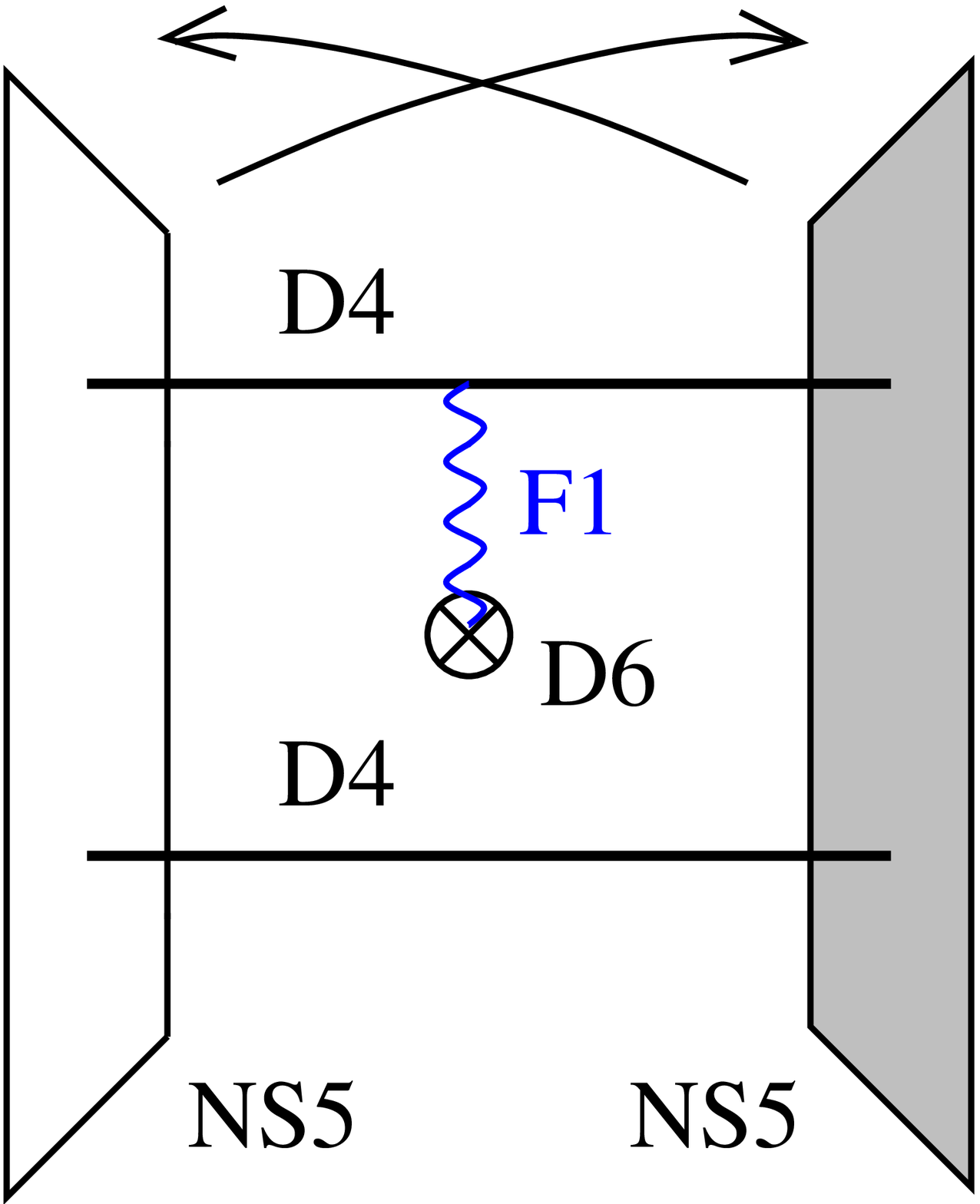}}
\end{picture}
\end{center}
}
$\Longrightarrow$
\parbox{4cm}{
\unitlength=.4mm
\begin{center}
\begin{picture}(100,80)
\epsfxsize=3cm
\put(10,-5){\epsfbox{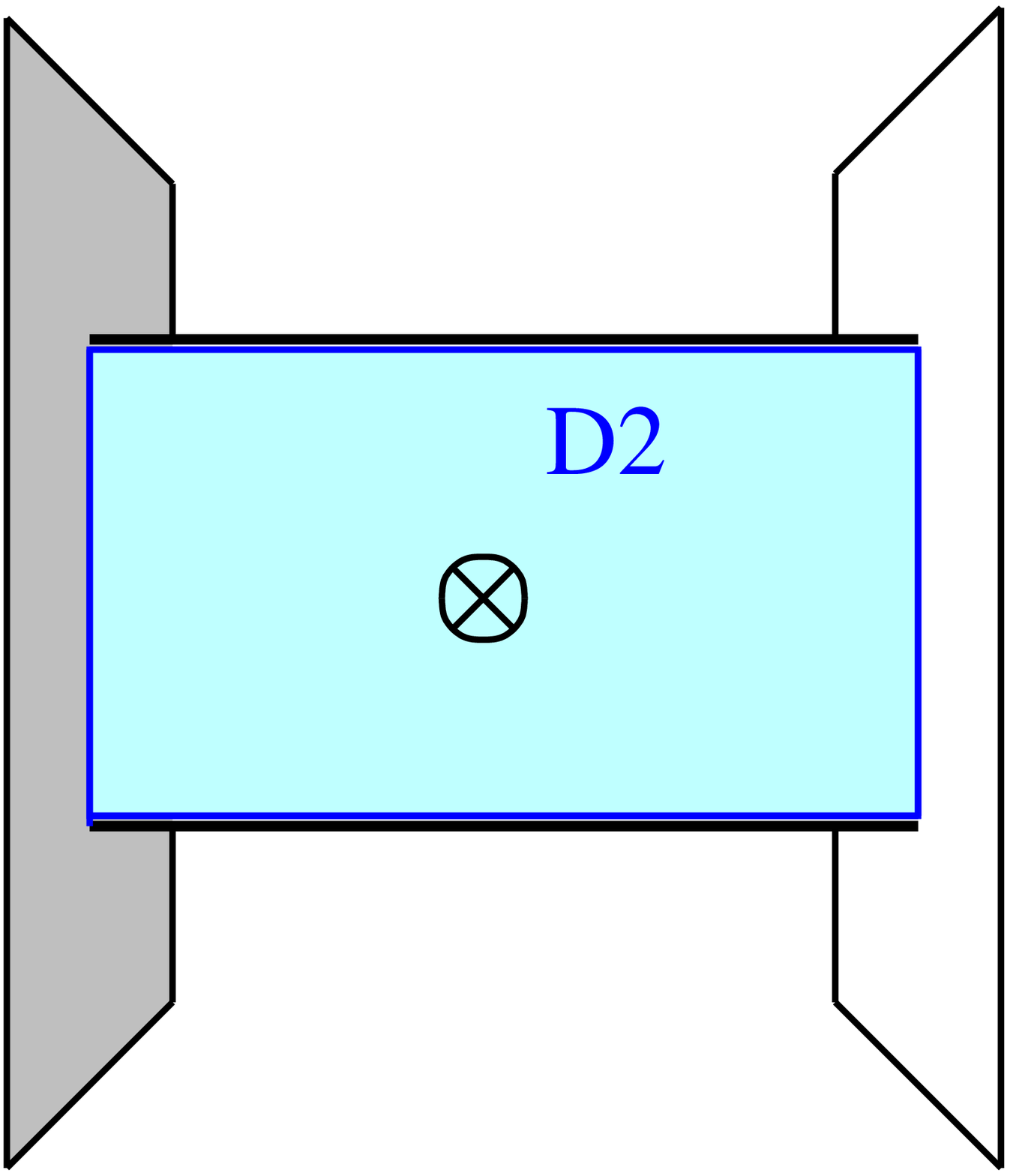}}
\end{picture}
\end{center}
}
\end{center}
\vspace{2ex}
\begin{center}
\refstepcounter{figure}
\label{fig13}
{Fig. \thefigure}
\end{center}
\vspace{2ex}

\subsection{Geodesics}
\label{geodesic}

In Ref.~\citen{KLMVW}, it is argued
that the BPS states are represented by
the geodesics on $\Sigma$ with the metric $ds^2=|\lambda|^2$,
where $\lambda$ is the Seiberg-Witten 1-form.
However, it was found in Ref.~\citen{M} that several
examples of the BPS states in MQCD do not correspond to the geodesics.
In this subsection, we first resolve this discrepancy
and then apply the technique in Ref.~\citen{KLMVW,BS} and \citen{R}
 to our situation.
Related arguments can also be found in a recent 
interesting paper Ref.~\citen{MNS}.

The mass of a particle in MQCD is proportional to the area of
the membrane:
\beq
{\rm mass}\propto\int_{\Sigma'}{\rm (vol)}
=\int_{\Sigma'}\sqrt{K^2+|\Omega|^2}.
\eeq
Here (vol) is the volume form on $\Sigma'$, 
and $K$ and $\Omega$ are the K\"ahler form (\ref{kahler}) and
 holomorphic 2-form
(\ref{omega}) on $Q$ pulled back to $\Sigma'$.\cite{FS,HY,M}
As shown in Ref.~\citen{FS,HY,M}, the BPS condition implies
that $\Omega$ has a constant phase and $K=0$ on $\Sigma'$, which is
equivalent to the condition that $\Sigma'$ is holomorphically embedded
in $Q$ with respect to a complex structure which is orthogonal to the
original one. 
Therefore,
\beq
{\rm mass}\propto\int_{\Sigma'}|\,\Omega\,|
=\left|\,\int_{\Sigma'}\Omega\,\right|
=\left|\,\int_{\partial\Sigma'}\lambda\,\right|,
\label{mass}
\eeq
where $\lambda\equiv 2\,\hat v \frac{d Y}{Y}$
 is a 1-form satisfying $\Omega=d\lambda$. The pull back of $\lambda$
to $\Sigma$ is proportional to the Seiberg-Witten 1-form.
Note that if $\Sigma$ passes through the points with
$Y=0$, we should also include
integration around these points.
Hence  $\partial\Sigma'$ in (\ref{mass}) may have components which
are not included in $\Sigma$.
Note also that at the point with $Y=0$,
(\ref{YZ}) implies $\hat v=m_j$ for some $j$, and thus
the integration of $\lambda$ around these points
is proportional to $m_j$.\cite{M}

The statement that $\partial\Sigma'$ is a geodesic on $\Sigma$
with the metric $ds^2=|\lambda|^2$ is equivalent to the
statement that $\lambda$ has a constant phase on $\partial\Sigma'$.
However, (\ref{mass}) does not necessarily imply
$|\int\lambda|=\int|\lambda|$.
The problem is that there is an ambiguity in the definition
of $\lambda$. We can use $\lambda+d f$ instead of $\lambda$ 
without changing the entire story.
Here $f$ is a function on $Q$ which is well-defined at least locally.
Let us show that $\partial\Sigma'$ is indeed
a geodesic on $\Sigma$ with the metric $ds^2=|\lambda+d f|^2$ for
some function $f$. 
Since $\Omega$ has a constant phase, we can assume
$\real \Omega=0$ on $\Sigma'$. Then
the relations $0=\real\Omega=d \real \lambda$ imply
that there is a real function $f_R$ which satisfies
$\real\lambda=-d f_R$ on $\Sigma'$.
Therefore if we choose $f$ such that $\real f=f_R$
on $\Sigma'$, $\lambda+d f$ has a constant phase on $\Sigma'$.

As an example, let us consider $\N=2$ $SU(2)$ SYM with $4$
massless flavors and consider the case
in which  $\Sigma'$ is homeomorphic to a disk.  
We define a 1-form $\lambda_f$ on $\Sigma$ as
\beq
\lambda_f=\lambda_{f\,i}\, dx^i\equiv\lambda+d f,
\eeq
where $f$ is a function on $\Sigma$.
Taking a suitable affine parameter $t$,
a geodesic on $\Sigma$ with the metric $ds^2=|\lambda_f|^2$
satisfies the geodesic equation
\beq
\lambda_{f\,i}\,\frac{dx^i}{dt}=\alpha,
\eeq
where $\alpha$ is a constant.
Integrating this equation, we obtain
\beq
\int^{x(t)}_{x(0)}\lambda+f(x(t))-f(x(0))=\alpha \,t.
\label{Jac}
\eeq

Now, explicit calculation shows that $\lambda$ is proportional to
a holomorphic 1-form on $\Sigma$ up to an exact 1-form,
and so, redefining $f$, we can set $\lambda\propto \frac{dx}{y}$.
Here we define the map
\beq
\varphi_f :\Sigma\ra J(\Sigma)
\eeq
by choosing a point $P_0\in\Sigma$ and setting
\beq
\varphi_f(P)=\int_{P_0}^P\lambda+f(P)-f(P_0).
\label{varphif}
\eeq
Here, $J(\Sigma)$ is the Jacobian of $\Sigma$ which is defined as
$J(\Sigma)\equiv\C/\Gamma$, where 
$\Gamma\equiv\Z\int_{\alpha}\lambda\oplus\Z\int_{\beta}\lambda$
is a lattice in $\C$.
Equation (\ref{Jac}) implies that the image of the geodesic
under the map $\varphi_f$ is a straight line on $J(\Sigma)$

Let us first consider the $f=0$ case.
A closed straight line on $J(\Sigma)$ is 
a line winding $p$ times around
the $\alpha$ cycle and $q$ times around the $\beta$ cycle,
where $p$ and $q$ are relatively prime integers.
 As is well known (see for example Ref.~\citen{FK}), 
the map $\varphi_0$ is biholomorphic,
 and thus all the $(p,q)$ lines
correspond to geodesics on $\Sigma$.
This fact strongly suggests that there are hypermultiplets
with $(n_e,n_m)=(p,q)$ charge,
and moreover, that
there are no hypermultiplets carrying $(np,nq)$ charge with
$n\ge 2$.
It is clear that we cannot find further states, even if we take
into account the function $f$ in (\ref{varphif}).
This result is consistent with the prediction given in Ref.~\citen{SW},
that the BPS spectrum is invariant under the duality group $SL(2,\Z)$.
Note that this result has already been derived using different methods
in Ref.~\citen{GMS,Fay}.

If we naively generalize this argument to the case with $N_c>2$,
the Jacobian $J(\Sigma)$ is replaced by $\C/\Gamma$, where
$\Gamma=(\oplus_i\int_{\alpha_i}\lambda)
\oplus(\oplus_i\int_{\beta_i}\lambda$).
However, $\Gamma$ is in general a dense set in $\C$, and $\C/\Gamma$ is
no longer a manifold. We do not know how to resolve this problem, 
and leave it for future study.

\section{Discussions}
\label{sect5}
As we have emphasized,
MQCD is a democratic theory in which
there is no discrimination between elementary particles and
solitons.
This is why we can treat the electric and magnetic theories
simultaneously, and duality becomes manifest.
This is quite analogous to the S-duality in type IIB string theory,
which can be understood as a redefinition of a symplectic basis
of $T^2$ in M-theory \cite{Sch,A}.

On the other hand,
in order to describe the theory in field theory language,
namely, in order to write down the Lagrangian,
we must pick up the elementary particles,
which dominate in the weak coupling limit, from the particle spectrum.
There may be another choice of elementary particles which dominate in
another limit of the couplings.
If we choose another set of particles as the elementary particles,
the Lagrangian will become totally different from the original one.
This is the reason that
 duality in field theory is so mysterious and difficult to 
prove exactly.
Note that the brane exchange in Ref.~\citen{HW,EGK,EGKRS,NOYY} 
is not a duality transformation,
but a prescription to find another set of the elementary particles,
moving the gauge coupling from weak to strong.

We have argued in the $\N=2$ duality
that the elementary particles of the magnetic theory
appear as solitons in the electric theory.
Since Seiberg's $\N=1$ duality can be obtain as a deformation of $\N=2$
duality, we can extrapolate to conjecture
that the magnetic theory in Seiberg's duality
is also a theory of solitons, as suggested in Seiberg's paper \cite{S}.
We did not examine the spectrum of the $\N=1$ theory,
but at least in principle, we can analyze
it finding stable minimal surfaces.
It would be interesting to extend our arguments to non-BPS states
(see Ref.~\citen{W1}).

In this paper,
MQCD is defined as an effective field theory on an M5-brane world volume.
Hence, we must abandon the democracy once 
we write down the effective Lagrangian.
Moreover, we do not really know whether or not it is a consistent theory.
Therefore it is necessary to construct a microscopic definition of MQCD,
which may be easier to construct than that of M-theory.
We think that the construction of MQCD might be a first step to
go beyond the Lagrangian description of particle physics.

\section*{Acknowledgements}
We would like to thank our colleagues at Kyoto University for valuable
discussions and encouragement. We are also grateful to Y.Yoshida and
S.Terashima for useful discussions.

\end{document}